\documentclass[11pt]{article}
\usepackage[margin=1in]{geometry}
\usepackage[T1]{fontenc}
\usepackage[utf8]{inputenc}
\usepackage{lmodern}
\usepackage{amsmath}
\usepackage{amssymb}
\usepackage{bm}
\usepackage{graphicx}
\usepackage{booktabs}
\usepackage{tabularx}
\usepackage{siunitx}
\usepackage{microtype}
\usepackage{xcolor}
\numberwithin{equation}{section}
\numberwithin{figure}{section}
\numberwithin{table}{section}
\usepackage{hyperref}
\usepackage[noabbrev,nameinlink]{cleveref}
\usepackage{float}
\usepackage{placeins}
\graphicspath{{figures/}}
\hypersetup{hypertexnames=false,colorlinks=true,linkcolor=blue!45!black,citecolor=blue!45!black,urlcolor=blue!45!black}
\crefname{figure}{Fig.}{Figs.}
\crefname{table}{Table}{Tables}
\crefname{section}{Sec.}{Secs.}
\DeclareMathOperator{\Cov}{Cov}
\DeclareMathOperator{\sgn}{sgn}
\DeclareMathOperator{\RMSE}{RMSE}
\DeclareMathOperator*{\argmin}{arg\,min}
\newcommand{\E}{\mathbb{E}}
\newcommand{\R}{\mathbb{R}}
\newcommand{\supp}{Supplementary Material}

\title{Breakdown of Axis Separability and Structured Inversion\\in Finite-Domain Normalized Position Measurements:\\A Quadrant Photodetector Case Study}
\author{Zhiqian Su\\Institute of Quantum Electronics, School of Electronics,\\Peking University, Beijing, China\\\texttt{zqsu@pku.org.cn}}
\date{}

\begin{document}
\maketitle

\begin{abstract}
Finite-domain normalized spatial measurements combine an incident field, geometric support, local response, and channel weighting. These operations can break an otherwise separable response, so accurate single-axis calibration alone does not establish separability of a multidimensional measurement. We represent normalized readouts as channel-weight expectations under a parameter-dependent effective measure. For fixed reception and channel operators, local sensitivity is the covariance between the channel weight and the source score. This representation distinguishes axis separability, parameter mixing, and local recoverability, and applies at the appropriate detector level to continuous-electrode position-sensitive detectors, Shack--Hartmann and pyramid wavefront sensors, back-focal-plane split detection, and finite pixel arrays.

A quadrant photodetector (QPD) provides an analytically tractable case. Starting from the complete two-dimensional power integrals, we identify the spatial-moment origin of its lowest-order cross term. Axis calibration and system symmetry then constrain the inverse map, giving Axis-Anchored Cross-Residual Inversion (ACRI). Numerical tests of this QPD realization support the sensitivity relations and show that the constrained cross correction substantially reduces the off-axis bias retained by single-axis inversion. The results illustrate how the general measurement representation guides inverse construction, with performance bounded by the working domain, calibration state, and preserved symmetries.
\end{abstract}

\section{Introduction}\label{sec:introduction}

A quadrant photodetector (QPD) estimates spot position from normalized differences of quadrant powers and is used in laser tracking, free-space optical communication, and autocollimation measurement~\cite{Manojlovic2011,Li2019,Diao2022}. For a circularly symmetric Gaussian spot on an ideal infinite receiving plane, the two differential channels each obey a one-dimensional error-function relation and can therefore be inverted independently.

Real devices have finite sensitive surfaces and interquadrant dead zones; the spot shape and reception response can also depart from the ideal model. Existing studies have improved positioning through models of nonuniform spots and blind areas, nonlinear compensation, range extension, and data-driven calibration~\cite{Zhang2018,Wu2015,Zhang2019,Wang2020,Qiu2022}. One common class retains independent axis-wise functions and represents nonlinearity by modified error functions, polynomials, or other fitted curves~\cite{Wu2015,Wang2020,ZhangW2019}. Axis-wise modeling is not universal: related optical angle sensors also use two-dimensional calibration surfaces to compensate interaxis coupling~\cite{Ran2025}.

For a finite circular QPD, Manojlovi{\'c} quantified sensitivity, interaxis crosstalk, and their dependence on the spot-to-detector size ratio~\cite{Manojlovic2011}. The question here is not whether crosstalk exists, but a broader structural one: under which conditions does a finite-domain normalized measurement preserve or break axis separability, and does the resulting parameter mixing still permit local recovery?

Finite reception does more than compress the dynamic range or change single-axis nonlinearity. If the incident spatial distribution and the complete effective reception domain do not retain a compatible separable structure, the orthogonal-direction contribution does not cancel under normalization, and one channel can depend on multiple position parameters. This off-axis coupling can vanish exactly on symmetry axes, so even highly accurate single-axis calibration cannot establish separability of the full two-dimensional map. At the same time, parameter coupling does not imply loss of information: the joint map may remain locally invertible and stably recoverable. Axis separability, parameter mixing, and local recoverability must therefore be treated as distinct questions.

We formulate finite-domain normalized spatial measurements as channel-weight expectations under a parameter-dependent effective measure. This representation combines geometric support, spatial response, and channel projection in a common local-sensitivity relation. Under fixed reception and channel operators, the covariance of a channel weight with a source score identifies how a parameter enters that channel. The finite circular QPD supplies an analytic example, including the spatial-moment origin of its lowest-order cross term. The same representation also accommodates continuous-electrode position-sensitive detectors, Shack--Hartmann and pyramid wavefront sensors, back-focal-plane split detection, and finite pixel arrays at their respective detector planes.

As an application of this representation, we use QPD axis calibration and system symmetry to construct Axis-Anchored Cross-Residual Inversion (ACRI). Known forward structure constrains the remaining two-dimensional inverse. Controlled numerical tests examine the sensitivity relations, the contribution of the cross residual, and the limits imposed by calibration mismatch and symmetry changes. They evaluate the QPD realization of the theory; the other devices establish the scope of its readout representation.

\Cref{sec:qpd} addresses axis separability under finite QPD reception. \Cref{sec:theory} develops the general theory and distinguishes parameter mixing from local recoverability. \Cref{sec:inverse} constructs ACRI for the QPD, \cref{sec:simulations} presents numerical tests and applicability limits, and \cref{sec:conclusion} concludes.

\section{QPD measurement and axis separability}\label{sec:qpd}

\subsection{Normalized quadrant readout and the axis-independent model}\label{sec:qpd-readout}

Position measurement with a QPD partitions the incident optical field into power integrals and forms normalized differences from the quadrant powers. Let the detector plane be the two-dimensional $xOy$ plane. A circularly symmetric Gaussian spot centered at $(x_0,y_0)$ has intensity
\begin{equation}\label{eq:gaussian}
I(x,y;x_0,y_0)=I_0\exp\!\left[-\frac{(x-x_0)^2+(y-y_0)^2}{\omega^2}\right].
\end{equation}
Here $I_0$ is the peak intensity, $\omega$ is the radius at which the intensity falls to $1/e$ of its peak, and $x_0,y_0$ are dimensional spot-center coordinates. The received power in each quadrant is the integral over its effective photosensitive region:
\begin{equation}\label{eq:qpd-readout}
\begin{aligned}
P_k&=\iint_{D_k} I(x,y;x_0,y_0)\,dx\,dy,\\
S_x&=\frac{(P_{\mathrm I}+P_{\mathrm{IV}})-(P_{\mathrm{II}}+P_{\mathrm{III}})}{P_{\mathrm I}+P_{\mathrm{II}}+P_{\mathrm{III}}+P_{\mathrm{IV}}},\\
S_y&=\frac{(P_{\mathrm I}+P_{\mathrm{II}})-(P_{\mathrm{III}}+P_{\mathrm{IV}})}{P_{\mathrm I}+P_{\mathrm{II}}+P_{\mathrm{III}}+P_{\mathrm{IV}}}.
\end{aligned}
\end{equation}

\begin{figure}[H]\centering
\includegraphics[width=\textwidth,height=.62\textheight,keepaspectratio]{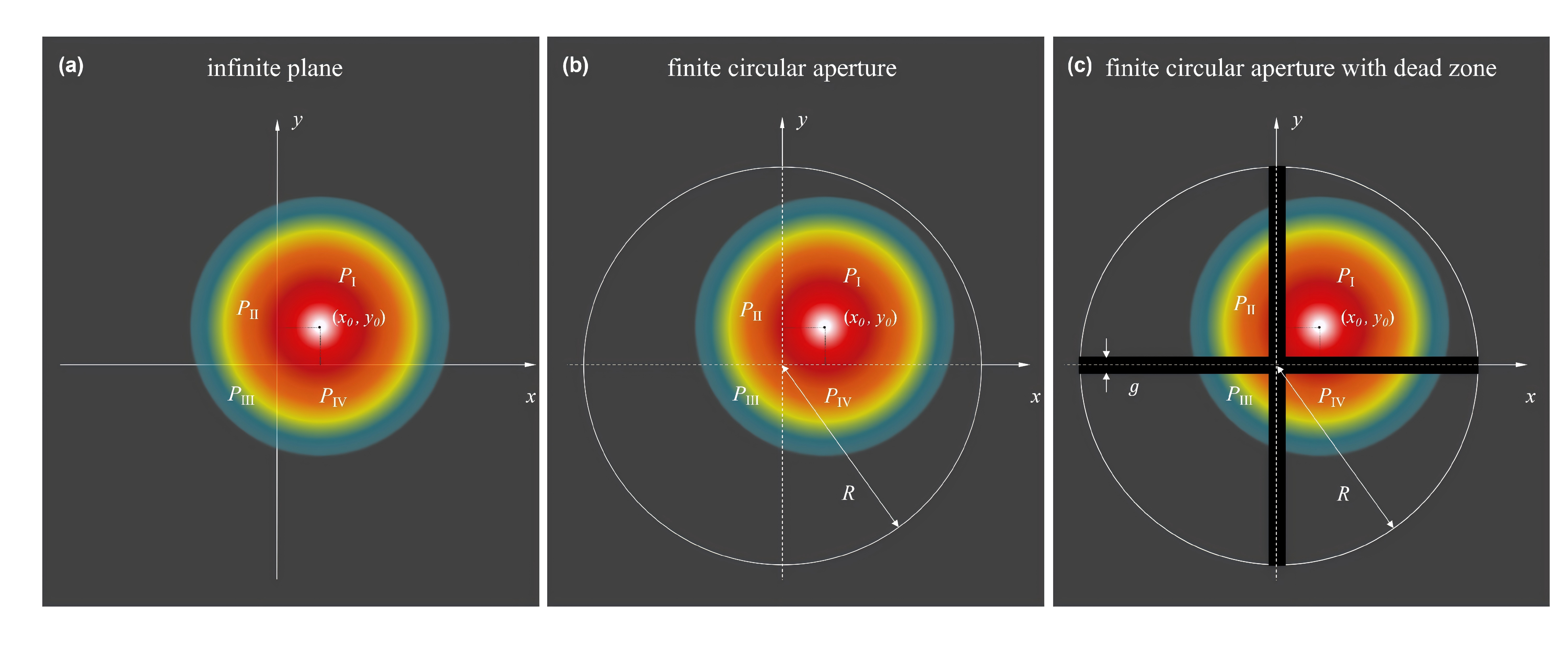}
\caption{QPD reception geometries considered in this work: (a) an infinite receiving plane, (b) a finite circular photosensitive surface, and (c) the same circular surface with a cross-shaped dead zone. $R$ is the detector radius, $g$ is the total dead-zone width, and $(x_0,y_0)$ denotes the spot center.}\label{fig:2-1}
\end{figure}

\cref{fig:2-1} highlights the geometric change from a separable receiving domain to a nonproduct one. For a circular photosensitive area, the admissible extent in one coordinate depends on the other coordinate. Adding the cross-shaped dead zone further removes part of the received field but leaves this nonproduct geometry unchanged.

\Cref{eq:qpd-readout} defines the full two-dimensional integrals, but obtaining a closed-form position inverse suited to real-time use is difficult when the finite boundary and dead zone are both present. QPD position models commonly begin with a simpler ideal limit and add corrections for device nonidealities~\cite{Wu2015,Wang2020}.

In the simplest case, the receiving plane is infinite, has no central dead zone, and the incident intensity factors in $x$ and $y$. The common orthogonal-direction integral cancels exactly in the normalized difference, giving the classical error-function response:
\begin{equation*}
S_x=\operatorname{erf}\!\left(\frac{x_0}{\omega}\right),\qquad
S_y=\operatorname{erf}\!\left(\frac{y_0}{\omega}\right).
\end{equation*}
This error-function relation is not an empirical fit; it follows directly from \eqref{eq:qpd-readout} under the stated ideal conditions. It gives the clearest axis-independent structure: the horizontal readout contains only $x_0$ and the vertical readout only $y_0$.

A real device first requires the central dead zone to be considered. Previous studies have introduced gap dimensions or effective response corrections into the ideal error-function model to describe changes in sensitivity and nonlinearity near quadrant boundaries~\cite{Wu2015,Wang2020}. A finite photosensitive surface further truncates the spot, narrows the usable working range, and changes the single-axis response; finite-aperture corrections, empirical calibration, lookup tables, and fitted functions have been used to extend position inversion~\cite{Wu2015,Wang2020}. These models differ in their functional forms and parametrizations, but commonly retain separate scalar forward relations for the two axes:
\begin{equation}\label{eq:axis-model}
S_x=F_x(x_0),\qquad S_y=F_y(y_0).
\end{equation}

Such one-dimensional corrections preserve the axis-independent form of \eqref{eq:axis-model}: each coordinate estimate uses only the differential signal from that axis~\cite{Wu2015,ZhangW2019,Wang2020}. This simplifies single-axis scans, fitting, and lookup, but the fit quality on an axis does not test whether the orthogonal coordinate influences the same channel.

This modeling route therefore leaves an assumption to examine. Axis independence on an ideal infinite plane follows from an exact product structure of the integration domain and separable optical field. If corrections for a dead zone or finite aperture still begin from two independent one-dimensional responses, they retain that structure in advance. Whether the actual finite two-dimensional receiving geometry permits the same decomposition cannot be decided from single-axis calibration accuracy alone. A finite aperture or gap may change not only the shapes of $F_x$ and $F_y$ but also the functional structure of the two-dimensional forward map. We therefore return to \eqref{eq:qpd-readout} without assuming \eqref{eq:axis-model}.

\subsection{Breakdown of axis separability and cross sensitivity}\label{sec:qpd-cross}

For the circular QPD in \cref{fig:2-1}(c), let $R$ be the photosensitive radius and $g$ the total width of the cross-shaped dead zone. The effective photosensitive region is
\begin{equation}\label{eq:qpd-domain}
D=\bigl\{(x,y):x^2+y^2\leq R^2,\ |x|\geq g/2,\ |y|\geq g/2\bigr\}.
\end{equation}

The circularly symmetric Gaussian in \eqref{eq:gaussian} factors as $I(x,y;x_0,y_0)=I_x(x;x_0)I_y(y;y_0)$, but factorization of the field does not imply factorization of the integrated readout. To expose the effect of the finite circular boundary, first set $g=0$. At fixed $x$, the allowed vertical interval is $-b(x)\leq y\leq b(x)$, where $b(x)=\sqrt{R^2-x^2}$. Let $\mathcal{F}_y(y;y_0)$ be any antiderivative of $I_y(y;y_0)$ with respect to $y$. The normalized horizontal difference is
\begin{equation}\label{eq:circular-integral}
\begin{aligned}
S_x(x_0,y_0)
&=\frac{\int_{-R}^{R}\sgn(x)I_x(x;x_0)
\left[\int_{-b(x)}^{b(x)}I_y(y;y_0)\,dy\right]dx}
{\int_{-R}^{R}I_x(x;x_0)
\left[\int_{-b(x)}^{b(x)}I_y(y;y_0)\,dy\right]dx}\\
&=\frac{\int_{-R}^{R}\sgn(x)I_x(x;x_0)
\left[\mathcal{F}_y(b(x);y_0)-\mathcal{F}_y(-b(x);y_0)\right]dx}
{\int_{-R}^{R}I_x(x;x_0)
\left[\mathcal{F}_y(b(x);y_0)-\mathcal{F}_y(-b(x);y_0)\right]dx}.
\end{aligned}
\end{equation}
The second equality merely evaluates the inner definite integral, but it exposes the coupling. Through $b(x)=\sqrt{R^2-x^2}$, the circular aperture places $x$ in the vertical integration limits. The bracketed factor consequently depends on both $x$ and $y_0$ and cannot be removed from the outer integral; the orthogonal coordinate $y_0$ remains in $S_x$.

Restoring the cross-shaped dead zone additionally excludes $|y|<g/2$ from the inner integral and $|x|<g/2$ from the outer integral. The integration ranges change, but the mechanism does not: when the complete effective reception domain cannot be written as two independent one-dimensional sets, the orthogonal-direction integral generally cannot cancel as a common factor. Special fields or integral cancellations can still be exceptions. A finite circular QPD therefore generally requires a genuinely two-dimensional forward map and may exhibit off-axis cross sensitivity:
\begin{equation}\label{eq:cross-forward}
S_x=F_x(x_0,y_0),\quad S_y=F_y(x_0,y_0),\qquad
\frac{\partial S_x}{\partial y_0}\not\equiv0,\quad
\frac{\partial S_y}{\partial x_0}\not\equiv0.
\end{equation}
``Not identically zero'' does not mean nonzero everywhere: geometric symmetry forces cross sensitivity to vanish at the origin and on certain symmetry axes.

\begin{figure}[H]\centering
\includegraphics[width=\textwidth,height=.66\textheight,keepaspectratio]{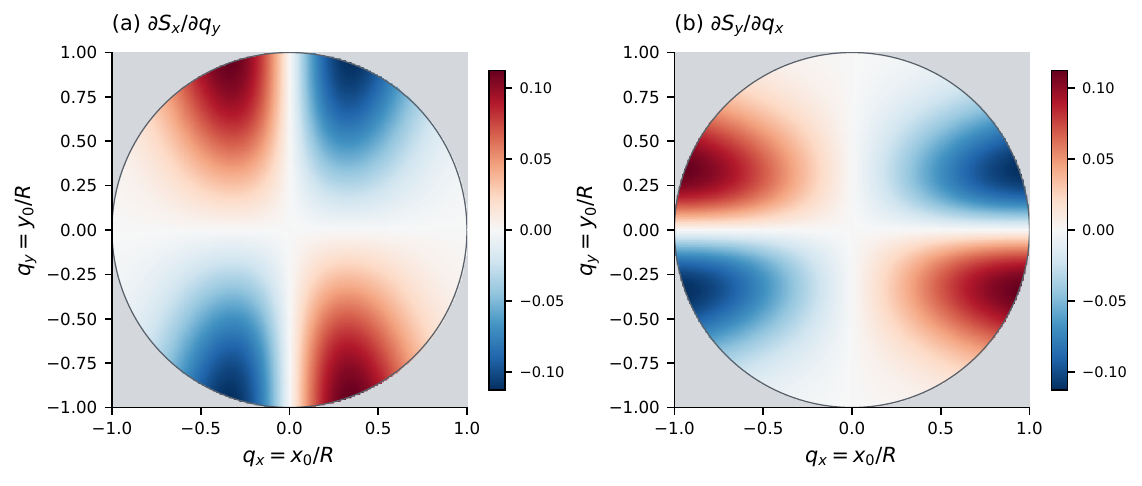}
\caption{Cross sensitivities over the full spot-center disk at $\rho=0.45$ and $g/R=0.032$: (a) $\partial S_x/\partial q_y$ and (b) $\partial S_y/\partial q_x$. The diverging scales identify positive and negative derivatives; the axes are symmetry-imposed zero lines. The dead zone restricts the receiving domain only; the spot-center domain remains the full disk.}\label{fig:2-2}
\end{figure}

\Cref{fig:2-2} shows zero cross sensitivity on the coordinate axes, as required by reflection symmetry, and nonzero response at general off-axis positions. The alternating signs across quadrants follow from the odd symmetry of the cross derivatives in both position coordinates, consistent with \eqref{eq:cross-forward} and \eqref{eq:forward-expansion} below. The plotted derivatives are with respect to normalized position.

For a centered QPD with identical quadrant responses, call the complete forward structure fourfold symmetric if it is invariant under $x\mapsto-x$, $y\mapsto-y$, and $x\leftrightarrow y$. Then $S_x$ is odd in $x_0$ and even in $y_0$, with the exchanged relations for $S_y$. The lowest-order expansion about the origin must have the form
\begin{equation}\label{eq:forward-expansion}
\begin{aligned}
S_x(x_0,y_0)&=A_{10}x_0+A_{30}x_0^3+A_{12}x_0y_0^2+O(5),\\
S_y(x_0,y_0)&=A_{10}y_0+A_{30}y_0^3+A_{12}y_0x_0^2+O(5).
\end{aligned}
\end{equation}
Here $A_{10}$ is the linear on-axis sensitivity near the origin, $A_{30}$ the lowest-order on-axis nonlinear coefficient, and $A_{12}$ the lowest-order off-axis cross-coupling coefficient. The coefficients are shared by the two axes because of $x\leftrightarrow y$ symmetry. The notation $O(5)$ denotes symmetry-allowed terms of total degree at least five in $(x_0,y_0)$.

\Cref{eq:forward-expansion} also explains why single-axis calibration cannot identify this coupling. For example, $A_{12}x_0y_0^2$ vanishes along the entire horizontal axis $y_0=0$. Even exact calibration of the one-dimensional response $S_x(x_0,0)$ therefore cannot establish whether $A_{12}$ vanishes. Single-axis data determine the on-axis response and its nonlinearity, but do not prove separability of the full two-dimensional map.

Conversely, finite reception does not necessarily cause interaxis coupling. If the \emph{complete} effective reception domain---including outer boundaries, dead zones, occlusions, and other geometric restrictions---is a product $D=D_x\times D_y$, and the incident field also factors as $I(x,y;x_0,y_0)=I_x(x;x_0)I_y(y;y_0)$, then the horizontal readout factors exactly:
\begin{equation}\label{eq:product-factorization}
\begin{aligned}
S_x&=\frac{\left[\int_{D_x}\sgn(x)I_x(x;x_0)\,dx\right]
\left[\int_{D_y}I_y(y;y_0)\,dy\right]}
{\left[\int_{D_x}I_x(x;x_0)\,dx\right]
\left[\int_{D_y}I_y(y;y_0)\,dy\right]}\\
&=\frac{\int_{D_x}\sgn(x)I_x(x;x_0)\,dx}
{\int_{D_x}I_x(x;x_0)\,dx}.
\end{aligned}
\end{equation}
The common vertical factor cancels, so $S_x$ depends only on $x_0$; the same reasoning applies to $S_y$.

Here ``product domain'' refers to the final effective domain after all reception restrictions have been combined, not only to the outer detector boundary. An ideal cross-shaped dead zone itself has a product structure: one may take $D_x=\{x\in\R:|x|\geq g/2\}$ and $D_y=\{y\in\R:|y|\geq g/2\}$. It therefore does not by itself break axis factorization on an infinite plane. An axis-aligned rectangular boundary combined with that gap can also retain a product structure. By contrast, a circular outer boundary makes the allowed $y$ range depend on $x$, so its complete domain with the gap is generally not a product.

Thus, a product effective domain together with a separable field gives a set of sufficient, but not necessary, conditions for axis separability. Some nonproduct regions may still be separable through special symmetry or integral cancellation.

\begin{figure}[H]\centering
\includegraphics[width=\textwidth,height=.72\textheight,keepaspectratio]{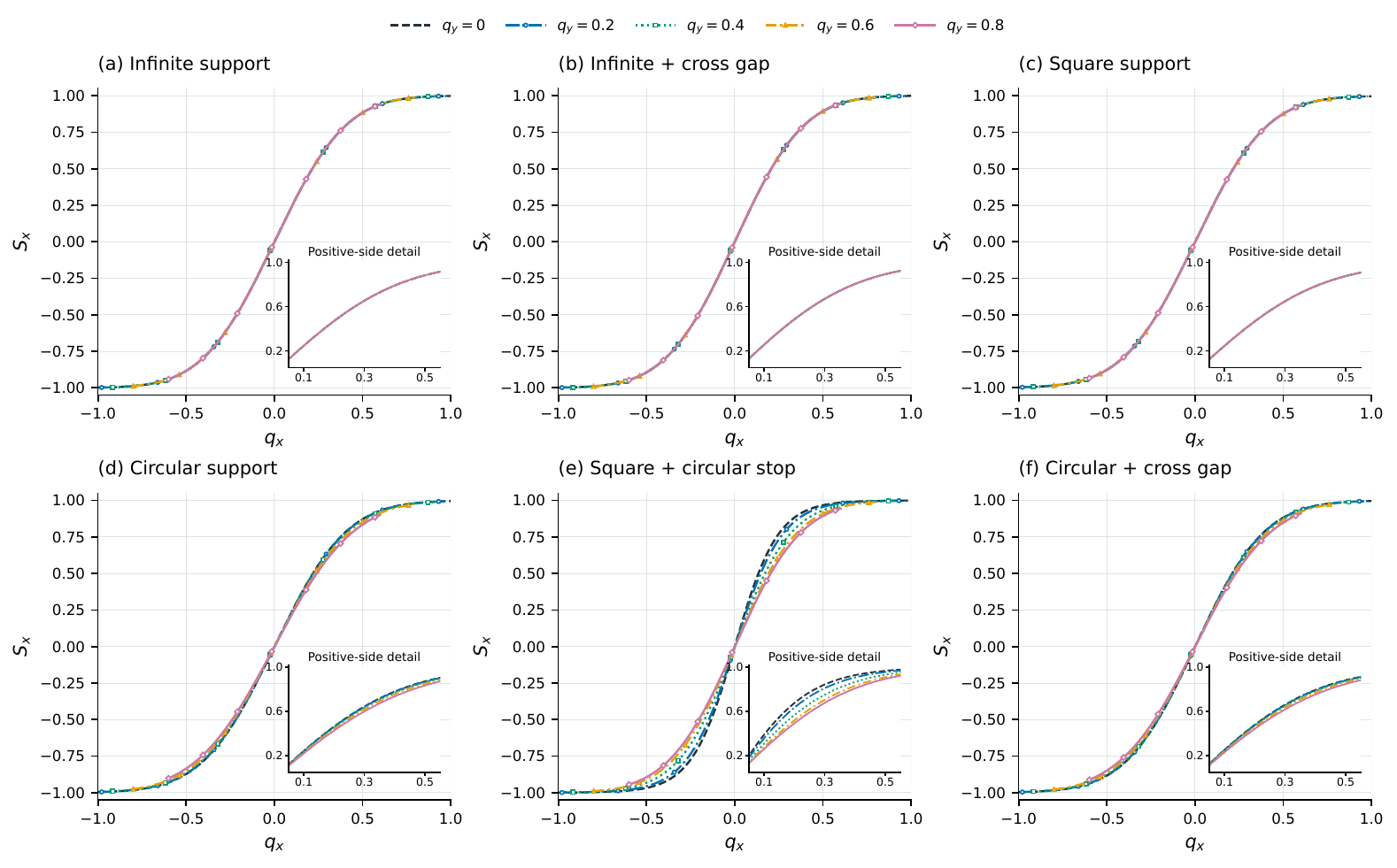}
\caption{Transverse response $S_x(q_x,q_y)$ for six receiving geometries at $\rho=0.45$: (a) infinite plane, (b) infinite plane with a cross gap, (c) square support, (d) circular support, (e) square support with an internal circular stop, and (f) circular support with a cross gap. Where present, $g/R=0.032$ and the circular stop has radius $0.5R$. Line styles identify $q_y$. Main axes retain the full signed-position response; insets enlarge the same positive-side data ($0.05\leq q_x\leq0.55$) without rescaling the response. Curves coincide for product domains (a--c) but separate for the nonproduct domains (d--f).}\label{fig:2-3}
\end{figure}

\Cref{fig:2-3} displays the response-level counterpart of \eqref{eq:circular-integral} and \eqref{eq:product-factorization}. For both the infinite plane and the rectangular product domain, horizontal response curves at different $y_0$ coincide despite their different reception conditions. A circular aperture or a nonproduct internal occlusion instead separates the $S_x$ curves at different off-axis positions. Separability is determined not by the labels ``finite aperture'' or ``dead zone'' alone but by the integral structure formed jointly by the field and the complete effective reception domain. The next section extends this geometric mechanism to general finite-domain normalized spatial measurements.

\section{General theory of finite-domain normalized position measurements}\label{sec:theory}

\subsection{Unified representation and effective measure}\label{sec:effective-measure}

\Cref{sec:qpd} showed that a two-dimensional QPD readout is not a direct sample of the incident field. It is a finite-dimensional projection formed after the continuous spatial distribution has been restricted by a finite sensitive domain, spatially weighted, and normalized by a common received signal. Apertures, gaps, and related structures therefore change both total received power and the parameter dependence of the normalized readout. This measurement chain is not specific to the QPD and can be expressed for a class of finite-domain normalized spatial measurements.

\begin{figure}[H]\centering
\includegraphics[width=\textwidth,height=.60\textheight,keepaspectratio]{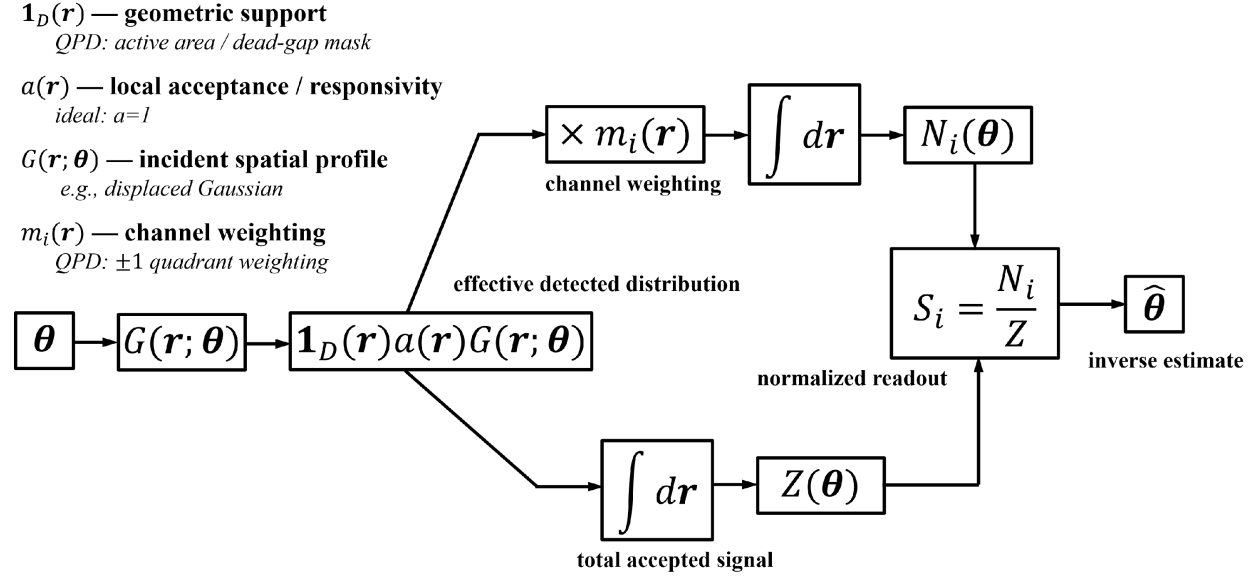}
\caption{Measurement chain for a finite-domain normalized spatial readout. The parameter $\bm\theta$ determines $G(\bm r;\bm\theta)$. Geometric support $\mathbf{1}_D$ and local response $a$ form the received field, which branches into a channel-weighted numerator $N_i$ and a common total signal $Z$. Their ratio is $S_i$, from which an inverse estimate may be formed. The QPD annotations are examples of the general quantities.}\label{fig:3-1}
\end{figure}

Let $\bm r\in\R^d$ denote detector-space position and $\bm\theta\in\Theta\subset\R^p$ the parameters to estimate. Write the parameter-dependent nonnegative spatial distribution as $G(\bm r;\bm\theta)$. For a planar QPD, $\bm r=(x,y)^{\mathsf T}$; for spot-position measurement, $\bm\theta=(x_0,y_0)^{\mathsf T}$ and $G$ may be a translated Gaussian intensity. Let $D$ be the effective reception domain and $\mathbf{1}_D(\bm r)$ its indicator. The nonnegative $a(\bm r)$ is local reception or response weight, while $m_i(\bm r)$ weights the received space for channel $i$. For an ideal QPD, $D$ combines the finite sensitive surface and central gap, $a(\bm r)=1$, and $m_i$ supplies the positive and negative quadrant weights for horizontal or vertical differences. We assume below that $D$, $a$, and $m_i$ are fixed with respect to the parameters being differentiated and that differentiation can be interchanged with integration under the stated regularity conditions. \Cref{fig:3-1} locates these quantities in the general measurement chain.

\Cref{fig:3-1} separates the source field, effective reception distribution, and channel projection. Geometry and local response first determine which spatial components are received; channel weights then integrate those components; finally, a common total signal normalizes the result. The same source field can therefore acquire different parameter dependence under different reception structures.

Let $N_i(\bm\theta)$ be the weighted integral for channel $i$ and $Z(\bm\theta)$ the total integral of the same effective received field. The normalized finite-domain readout is
\begin{equation}\label{eq:general-readout}
\begin{aligned}
N_i(\bm\theta)&=\int_{\R^d}m_i(\bm r)\mathbf{1}_D(\bm r)a(\bm r)G(\bm r;\bm\theta)\,d\bm r,\\
Z(\bm\theta)&=\int_{\R^d}\mathbf{1}_D(\bm r)a(\bm r)G(\bm r;\bm\theta)\,d\bm r,\\
S_i(\bm\theta)&=\frac{N_i(\bm\theta)}{Z(\bm\theta)}.
\end{aligned}
\end{equation}
For the uniform ideal QPD of \cref{sec:qpd}, the horizontal and vertical normalized differences correspond to
\begin{equation}\label{eq:qpd-weights}
m_x(\bm r)=\sgn(x),\qquad m_y(\bm r)=\sgn(y).
\end{equation}
The reception weight and channel weight have distinct roles. The product $\mathbf{1}_D a$ determines which spatial signal enters the measurement and with what local response; $m_i$ determines how the already received signal is projected onto a particular channel. Both are spatial functions, but they act at different stages of the forward measurement chain.

Normalize the actually received spatial distribution to define a parameter-dependent effective probability measure:
\begin{equation}\label{eq:effective-measure}
d\mu_{\bm\theta}(\bm r)=\frac{\mathbf{1}_D(\bm r)a(\bm r)G(\bm r;\bm\theta)}{Z(\bm\theta)}\,d\bm r,
\qquad \int_{\R^d}d\mu_{\bm\theta}(\bm r)=1.
\end{equation}
Then \eqref{eq:general-readout} becomes
\begin{equation}\label{eq:channel-expectation}
S_i(\bm\theta)=\int_{\R^d}m_i(\bm r)\,d\mu_{\bm\theta}(\bm r)
=\E_{\mu_{\bm\theta}}[m_i].
\end{equation}

\Cref{eq:channel-expectation} is the basic representation used below. The channel is not projected directly from the original field $G$: geometric support and local response first create the parameter-dependent effective measure $\mu_{\bm\theta}$, under which the channel weight is averaged. Apertures, dead zones, occlusions, and spatially nonuniform response are therefore components of the forward measurement operator, not error terms appended after an ideal readout. When $a(\bm r)=1$, $\mu_{\bm\theta}$ is the normalized restriction of $G$ to the effective receiving domain; nonuniform response further reshapes this distribution.

\begin{figure}[H]\centering
\includegraphics[width=\textwidth,height=.73\textheight,keepaspectratio]{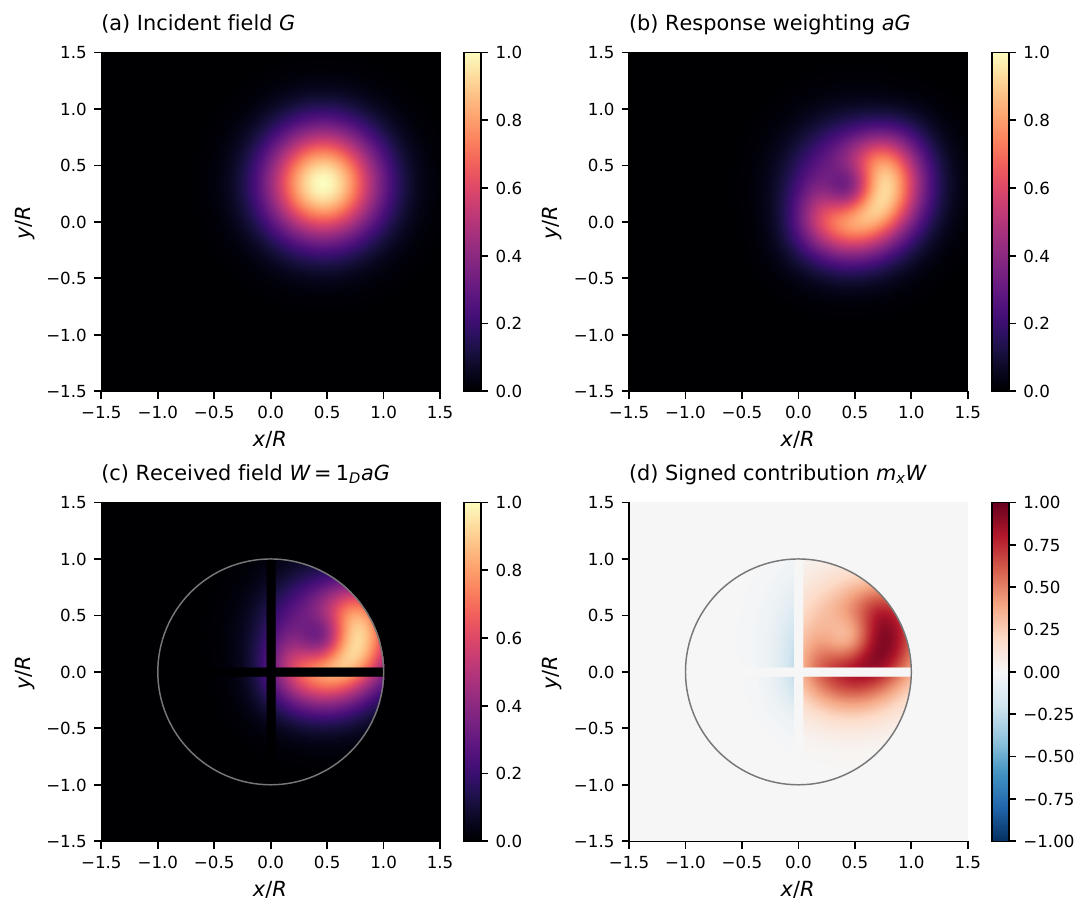}
\caption{Successive spatial weighting: (a) incident field $G$; (b) illustrative nonuniform response weighting $aG$; (c) received field $W=\mathbf{1}_D aG$ after circular-aperture and cross-gap masking; and (d) signed horizontal-channel contribution $m_xW$. The first three panels use a common absolute intensity scale; panel (d) uses a signed scale. The illustrative nonnegative $a$ is deliberately nonuniform, whereas nominal calculations use $a=1$. The QPD sign weight changes sign, not support or magnitude.}\label{fig:3-2}
\end{figure}

\Cref{fig:3-2} shows these layers rather than introducing a different forward model. The incident profile in (a) is reweighted, but not cut off, by the illustrative $a$ in (b). The aperture and dead zone remove support in (c). The signed colors in (d) are contributions to the horizontal numerator, not another change to received intensity. The nonuniformity is exaggerated for legibility; subsequent nominal calculations still set $a=1$.

Let $X(\bm r)=x$ and $Y(\bm r)=y$ denote coordinate functions under the effective measure. For the QPD, \eqref{eq:qpd-weights}--\eqref{eq:channel-expectation} give
\begin{equation}\label{eq:qpd-expectation}
S_x=\E_{\mu_{\bm\theta}}[\sgn X],\qquad
S_y=\E_{\mu_{\bm\theta}}[\sgn Y].
\end{equation}
The QPD differences are thus projections of the effective reception distribution onto two sign-weighted channels. More generally, $m_i$ need not be a continuous coordinate or quadrant sign function: it can represent coordinate moments, finite-partition indicators, pixel combinations, or other fixed spatial readout weights. \Cref{sec:systems} discusses the corresponding levels for continuous-electrode PSDs, Shack--Hartmann and pyramid wavefront sensors, back-focal-plane split detection, and pixel arrays. Their commonality is not one empirical readout formula, but the finite-domain normalized projection structure of \eqref{eq:channel-expectation}. In this representation, the interaxis coupling of \cref{sec:qpd} becomes a change in a channel expectation as $\mu_{\bm\theta}$ varies with the parameters.

\subsection{Score--channel covariance representation of parameter sensitivity}\label{sec:covariance}

\paragraph{General parameter sensitivity.}
\Cref{eq:channel-expectation} expresses a normalized readout as the expectation of a channel weight under a parameter-dependent measure. To describe the local spatial change induced by parameter $\theta_j$ in the forward field, define its \emph{source score} as
\begin{equation}\label{eq:source-score}
\ell_j(\bm r;\bm\theta)=\frac{\partial}{\partial\theta_j}\ln G(\bm r;\bm\theta).
\end{equation}
Here ``source score'' means only the local logarithmic response of the unnormalized forward field $G$ to a parameter; it is not automatically a statistical likelihood or Fisher score. Assume $0<Z<\infty$, $G>0$ almost everywhere on the effective support, finite required moments, and interchangeability of parameter differentiation and integration. Also assume that $D$, $a$, and $m_i$ do not depend on the differentiated parameter. Applying $\partial_{\theta_j}G=G\ell_j$ to the numerator and denominator of \eqref{eq:general-readout} gives the parameterized-expectation score-function identity~\cite{Mohamed2020}:
\begin{equation}\label{eq:covariance-identity}
\frac{\partial S_i}{\partial\theta_j}
=\E_{\mu_{\bm\theta}}[m_i\ell_j]
-\E_{\mu_{\bm\theta}}[m_i]\E_{\mu_{\bm\theta}}[\ell_j]
=\Cov_{\mu_{\bm\theta}}(m_i,\ell_j).
\end{equation}
This identity gives the local sensitivity of a finite-domain normalized measurement. Whether a parameter enters a channel depends on the covariance, under the effective measure, of the source-field change $\ell_j$ with the channel weight $m_i$. Apertures, gaps, occlusions, and nonuniform response need not appear explicitly in the derivative formula: they reshape $\mu_{\bm\theta}$ and hence the spatial projection. Cross sensitivity is part of the forward measurement structure, not a post-processing error added after a two-dimensional readout.

\paragraph{Two-dimensional local sensitivity of a QPD.}
For the circularly symmetric translated Gaussian used in \cref{sec:qpd}, the source scores for $x_0,y_0$ are
\begin{equation}\label{eq:gaussian-score}
\ell_{x_0}=\frac{2(x-x_0)}{\omega^2},\qquad
\ell_{y_0}=\frac{2(y-y_0)}{\omega^2}.
\end{equation}
Substituting \eqref{eq:gaussian-score} and the QPD weights $m_x=\sgn X$, $m_y=\sgn Y$ into \eqref{eq:covariance-identity}, the constants $x_0,y_0$ cancel in the covariances. The two-dimensional readout Jacobian is
\begin{equation}\label{eq:qpd-jacobian}
J=\frac{2}{\omega^2}
\begin{pmatrix}
\Cov_{\mu_{\bm\theta}}(\sgn X,X)&\Cov_{\mu_{\bm\theta}}(\sgn X,Y)\\
\Cov_{\mu_{\bm\theta}}(\sgn Y,X)&\Cov_{\mu_{\bm\theta}}(\sgn Y,Y)
\end{pmatrix}.
\end{equation}
The diagonal elements describe the local response of each differential channel to its own position coordinate, and the off-diagonal elements are cross sensitivities. For example, $J_{xy}=\partial S_x/\partial y_0$ is proportional to $\Cov_{\mu_{\bm\theta}}(\sgn X,Y)$. The off-axis coupling seen directly from integration limits in \cref{sec:qpd-cross} is therefore a local geometric criterion: if the effective measure after finite reception does not retain zero covariance between the horizontal sign channel and vertical coordinate, a vertical displacement enters the horizontal readout.

In \eqref{eq:qpd-jacobian}, $J=\partial(S_x,S_y)/\partial(x_0,y_0)$ is differentiated with respect to physical coordinates and has dimension inverse length. For $\bm q=(x_0/R,y_0/R)$, the dimensionless Jacobian is $J_q=RJ$. Position derivatives and singular values in the numerical figures use this latter convention.

\paragraph{From local sensitivity to the lowest-order cross coefficient.}
\Cref{eq:covariance-identity,eq:qpd-jacobian} hold at any operating point. For the centered fourfold-symmetric QPD of \cref{sec:qpd-cross}, the lowest-order cross coefficient near the origin can also be obtained analytically. Write the translated Gaussian as the centered field times an exponential tilt:
\begin{equation}\label{eq:exponential-tilt}
I(x,y;x_0,y_0)=I(x,y;0,0)
\exp\!\left(\frac{2xx_0+2yy_0-x_0^2-y_0^2}{\omega^2}\right).
\end{equation}
The final two terms do not depend on detector-plane position and cancel exactly between numerator and denominator of a normalized readout. Let $\mu_0\equiv\mu_{(0,0)}$, $A=2x_0/\omega^2$, and $B=2y_0/\omega^2$. Then
\begin{equation}\label{eq:tilted-response}
S_x(x_0,y_0)=
\frac{\E_{\mu_0}[\sgn(X)e^{AX+BY}]}
{\E_{\mu_0}[e^{AX+BY}]}.
\end{equation}
The centered field, finite reception geometry, and spatial response are all absorbed into the fixed measure $\mu_0$; displacement enters only through $A$ and $B$. This exponential-tilt representation makes the Taylor coefficients near the origin accessible through spatial moments under $\mu_0$, consistent with the standard moment relations for exponential families~\cite{WainwrightJordan2008}.

If the centered effective measure is mirror symmetric in both $x$ and $y$, odd moments vanish by symmetry. Retaining the moments needed for the cubic cross term gives
\begin{equation}\label{eq:moment-expansion}
\begin{aligned}
\E_{\mu_0}[\sgn(X)e^{AX+BY}]&=
A\E_{\mu_0}[|X|]
+\frac{A^3}{6}\E_{\mu_0}[|X|^3]
+\frac{AB^2}{2}\E_{\mu_0}[|X|Y^2]+\cdots,\\
\E_{\mu_0}[e^{AX+BY}]&=
1+\frac{A^2}{2}\E_{\mu_0}[X^2]
+\frac{B^2}{2}\E_{\mu_0}[Y^2]+\cdots.
\end{aligned}
\end{equation}
Expanding the ratio and comparing with $S_x=A_{10}x_0+A_{30}x_0^3+A_{12}x_0y_0^2+O(5)$ from \eqref{eq:forward-expansion} yields
\begin{equation}\label{eq:A12}
A_{12}=\frac{4}{\omega^6}
\left(\E_{\mu_0}[|X|Y^2]-\E_{\mu_0}[|X|]\E_{\mu_0}[Y^2]\right)
=\frac{4}{\omega^6}\Cov_{\mu_0}(|X|,Y^2).
\end{equation}
This is the spatial-moment origin of the lowest-order cross term. If the centered effective measure is a product, $|X|$ and $Y^2$ factor under it, their covariance is exactly zero, and $A_{12}=0$. For a finite circular aperture, the permitted vertical cross-section depends on $X$, so $|X|$ and $Y^2$ generally do not have that product relation and $A_{12}$ can be nonzero. More generally, even with a product outer geometry, a nonproduct spatial response $a(\bm r)$ can reshape $\mu_0$ and produce the same type of cross term.

For a uniform circular QPD with no gap, set $g=0$, $a(\bm r)=1$, and $D=\{x^2+y^2\leq R^2\}$. By coordinate-exchange symmetry, $R^3A_{12}$ here represents the same third-order mixed response as the $S_{yxx}$ defined by Manojlovi{\'c}~\cite{Manojlovic2011}. Evaluating \eqref{eq:A12} in this special case gives a coefficient that differs from the published expression in its Eq.~(37). Direct differentiation of the complete normalized two-dimensional forward integral and an independent high-accuracy integration--finite-difference calculation support the present expression; the linear sensitivity in its Eq.~(31) is reproduced. Details of this special-case comparison, including the closed form and two spot-size limits, are provided in the \supp.

The $x_0y_0^2$ term is therefore not an empirical correction introduced to fit an off-axis error. It is the lowest-order two-dimensional term allowed jointly by finite reception, a translated Gaussian, and mirror symmetry. It vanishes along the entire horizontal axis $y_0=0$; correspondingly, $\partial S_x/\partial y_0=2A_{12}x_0y_0+O(4)$ vanishes at the origin and on both axes by symmetry, while appearing at general off-axis positions. High-accuracy axis scans alone cannot establish separability of the full two-dimensional forward map. Continuing the same exponential-tilt and moment expansion gives higher symmetry-allowed terms whose coefficients are likewise determined by higher spatial moments under $\mu_0$, without prescribing an empirical cross polynomial.

\subsection{Separability, parameter mixing, and local recoverability}\label{sec:recoverability}

The previous subsections show from finite-domain integration and local sensitivity that cross response depends on the complete forward measurement structure. We now distinguish three concepts: coordinate separability, parameter mixing, and local recoverability. Consider two parameters $\bm\theta=(\theta_1,\theta_2)^{\mathsf T}$ and two readouts $\bm S=(S_1,S_2)^{\mathsf T}$. For the QPD, these are $(x_0,y_0)$ and $(S_x,S_y)$; in other systems, $\theta_1,\theta_2$ may be general unknown quantities such as local wavefront slopes, modal coefficients, or particle-displacement components. They need not be Cartesian or geometrically orthogonal.

\paragraph{1. Parameter-coordinate separability.}
For a given parametrization, the forward map is separable on a working domain if there exist scalar functions $F_1,F_2$ throughout that domain such that $S_1=F_1(\theta_1)$ and $S_2=F_2(\theta_2)$. For a continuously differentiable map this requires the cross derivatives to vanish throughout the domain:
\begin{equation}\label{eq:separability-derivatives}
\frac{\partial S_1}{\partial\theta_2}=0,\qquad
\frac{\partial S_2}{\partial\theta_1}=0.
\end{equation}
\Cref{sec:qpd-cross} gave a concrete sufficient set of conditions: a product complete effective domain, a source field separable in the two parameters, and compatible product forms of reception and channel weights make the common orthogonal factor cancel in the normalized readout. The condition concerns the complete forward operator, not simply whether the outer boundary is finite. By \eqref{eq:covariance-identity}, \eqref{eq:separability-derivatives} can also be read as sustained zero covariance of each cross-parameter source score with the other channel weight under the effective measure. A cross derivative that vanishes at one point or along a symmetry axis only indicates local symmetry cancellation, not separability over the working domain.

\paragraph{2. Parameter mixing.}
When \eqref{eq:separability-derivatives} fails, a channel depends on both parameters and the forward map exhibits parameter mixing. Define the local Jacobian
\begin{equation}\label{eq:general-jacobian}
J(\bm\theta)=\frac{\partial\bm S}{\partial\bm\theta}
=\begin{pmatrix}
\partial_{\theta_1}S_1&\partial_{\theta_2}S_1\\
\partial_{\theta_1}S_2&\partial_{\theta_2}S_2
\end{pmatrix}.
\end{equation}
Nonzero off-diagonal entries mean that the two parameters are no longer encoded independently by the two channels, but they do not alone imply loss of parameter information. For the QPD, $A_{12}x_0y_0^2$ is a concrete example: it disappears on an axis by symmetry and generates a cross derivative off axis. Axis separability and local recoverability impose different constraints. The former requires the relevant cross dependence to vanish, though a diagonal Jacobian may still have a zero diagonal entry. The latter requires the joint readout to distinguish two parameter directions locally; the Jacobian need not be diagonal. Neither property is simply a stronger or weaker version of the other.

\paragraph{3. Local recoverability and conditioning.}
If $\det J(\bm\theta)\neq0$ at an operating point, the inverse-function theorem gives a local inverse in a neighborhood of that point. Two parameters can thus in principle be recovered jointly despite nonzero cross sensitivity, provided the Jacobian remains full rank. Local invertibility does not guarantee stable inversion. For a small readout perturbation, define the linearized parameter increment $\delta\bm\theta_{\mathrm{lin}}=J^{-1}\delta\bm S$. Its amplification satisfies~\cite{Higham2002}
\begin{equation}\label{eq:singular-bound}
\|\delta\bm\theta_{\mathrm{lin}}\|_2
\leq\frac{1}{\sigma_{\min}(J)}\|\delta\bm S\|_2.
\end{equation}
Here $\sigma_{\min}(J)$ is the smallest singular value. It describes the weakest locally recoverable direction in the joint measurement: a sufficiently large value permits stable discrimination even with mixing, while a value approaching zero makes one parameter combination nearly indistinguishable in the readout and greatly amplifies noise or model error. \Cref{eq:singular-bound} bounds a linearized increment, not the exact global error under an arbitrary finite perturbation. Singular-value values depend on parameter units and scaling; conditioning comparisons must use the same parameter coordinates. QPD numerical results use $J_q$ throughout.

\begin{figure}[H]\centering
\includegraphics[width=\textwidth,height=.65\textheight,keepaspectratio]{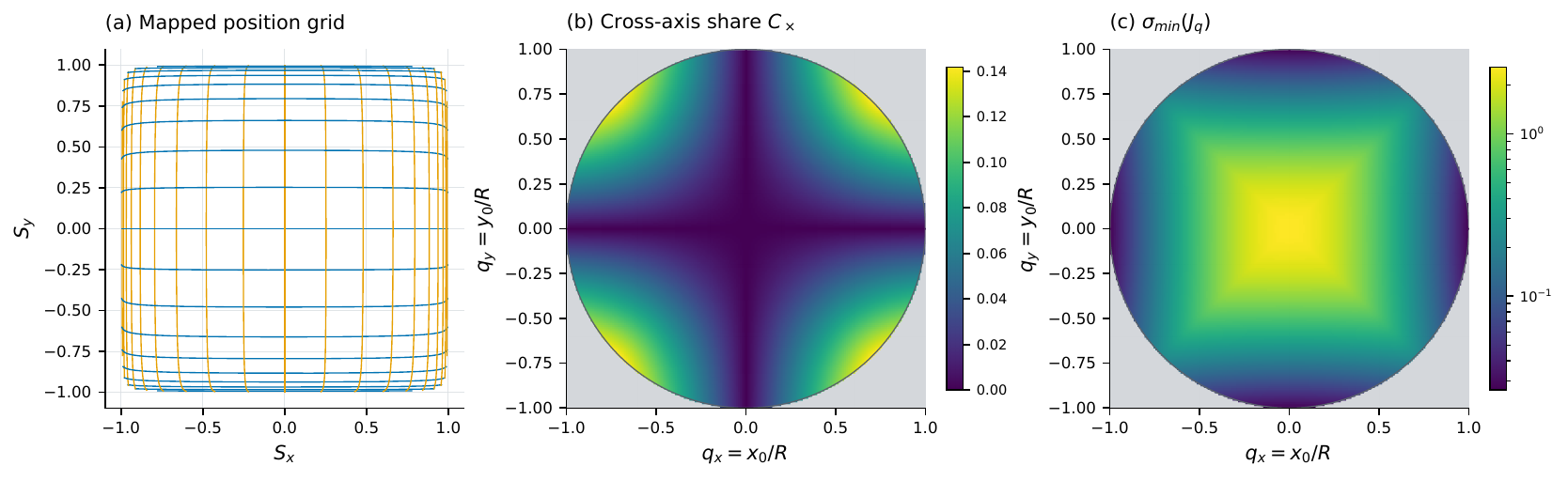}
\caption{Local mixing and recoverability over the full spot-center disk at $\rho=0.45$ and $g/R=0.032$. (a) A position grid mapped into $(S_x,S_y)$ readout coordinates; (b) the cross-axis share (Jacobian fraction) $C_\times$ defined in \eqref{eq:cross-fraction}; (c) $\sigma_{\min}(J_q)$ on a logarithmic color scale. Mixing and loss of local invertibility are distinct; these sampled maps neither imply a causal relation nor establish global injectivity.}\label{fig:3-3}
\end{figure}

On a working domain with connected coordinate sections, identically vanishing cross sensitivities correspond to axis-separated encoding. For mixed encoding, local inversion depends separately on the rank and smallest singular value of the joint Jacobian. To visualize the relative cross sensitivity in \cref{fig:3-3}, define
\begin{equation}\label{eq:cross-fraction}
C_\times(\bm q)=\frac{\sqrt{(J_q)_{xy}^{\,2}+(J_q)_{yx}^{\,2}}}{\|J_q\|_F},
\qquad \|J_q\|_F>0,
\end{equation}
where $J_q=\partial(S_x,S_y)/\partial(q_x,q_y)$ and $\|\cdot\|_F$ is the Frobenius norm. Thus $C_\times$ measures the relative magnitude of the off-diagonal entries in the stated normalized coordinates; conditioning is assessed separately by $\sigma_{\min}(J_q)$.

In \cref{fig:3-3}(a), the mapped coordinate grid bends away from a Cartesian grid, showing that the channels jointly encode the two positions. The cross-axis share in (b) is zero on the axes and increases off axis, while (c) shows the smallest singular value changing over the same disk yet remaining nonzero at the sampled positions. Cross response is therefore not equivalent to a loss of local recoverability. Mixing and conditioning may both change toward the edge, but they are not causally equivalent, and the figure does not prove global injectivity over the continuous domain. Full rank of a Jacobian ensures only local invertibility. A practical two-dimensional inverse still requires axis-response monotonicity, working-domain restrictions, and possible distant folds of the forward map to be considered. Structured inversion below is discussed only on a finite working domain constrained by those conditions.

\subsection{Representative measurement systems and scope of the framework}\label{sec:systems}

\begin{figure}[!htbp]\centering
\includegraphics[width=\textwidth,height=.77\textheight,keepaspectratio]{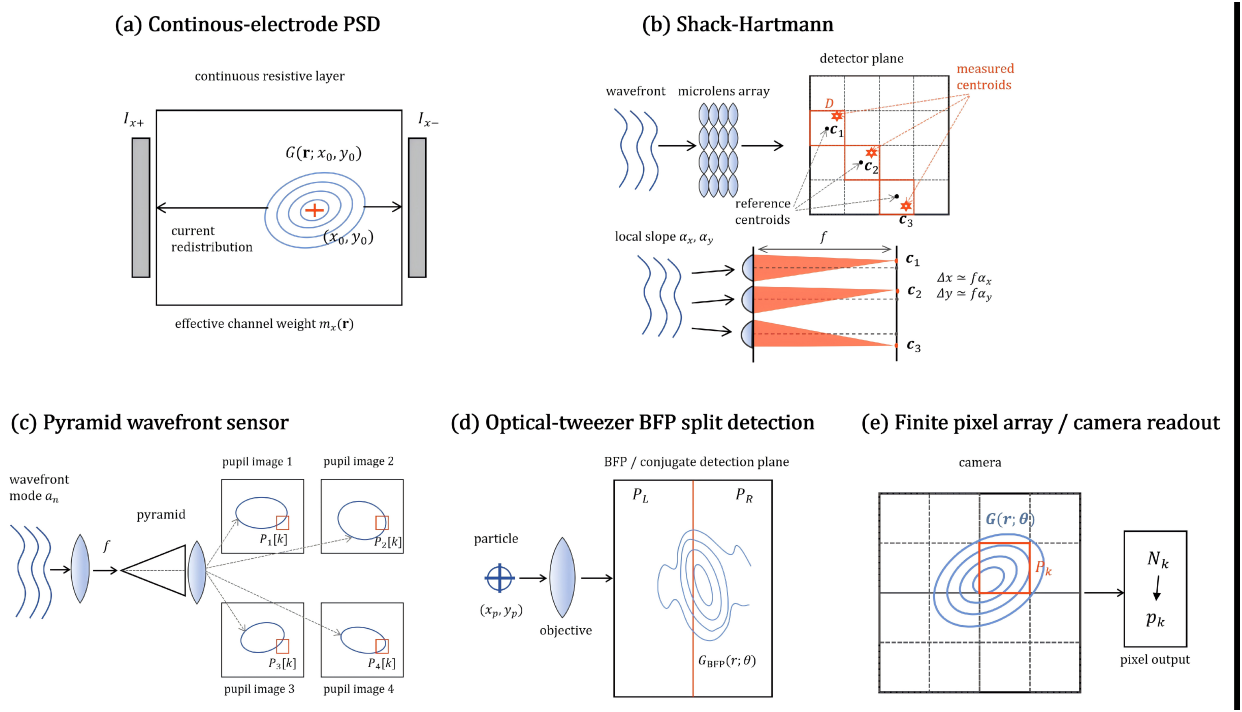}
\caption{Readout structures of five finite-domain normalized spatial measurements. (a) A continuous-electrode position-sensitive detector (PSD) redistributes photocurrent through a resistive layer toward edge electrodes. (b) A Shack--Hartmann sensor converts local wavefront slope into focal-spot centroid displacement through a microlens array. (c) A pyramid wavefront sensor (PWFS) creates multiple pupil images and forms differential channels from corresponding pixel combinations. (d) Back-focal-plane (BFP) split detection in optical tweezers senses particle-displacement-induced changes in the interference intensity at a conjugate detector plane. (e) A finite pixel array or camera discretizes the detector plane into local integration channels.}\label{fig:3-4}
\end{figure}

The preceding theory used the QPD as a canonical case, but \eqref{eq:channel-expectation} and \eqref{eq:covariance-identity} do not require quadrant detection. Continuous-electrode position-sensitive detectors, Shack--Hartmann sensors, pyramid wavefront sensors, back-focal-plane split detection in optical tweezers, and finite pixel arrays have different fields and channels. At a suitable readout level, however, each can be represented as a normalized projection of a finitely received distribution onto fixed spatial weights. \Cref{fig:3-4} sketches these five measurement processes.

The five systems are mapped onto the same framework according to their actual reception planes and readout channels; they do not share an optical-field model. PSD and local centroid readouts use continuous weights, split detection uses sign weights, and pixel arrays use local indicator functions. The source fields for the PWFS and BFP cases must include the intensity after propagation and interference. For any specific device, identifying the parameter-dependent detector-plane intensity $G(\bm r;\bm\theta)$, actual reception structure, and weights $m_i(\bm r)$ permits local parameter sensitivity to be evaluated using \eqref{eq:covariance-identity}.

\paragraph{Continuous-electrode position-sensitive detector (PSD).}
A continuous-electrode PSD redistributes photocurrent generated in a continuous photosensitive layer through a resistive network to edge electrodes; terminal current sums and differences estimate spot position. Let $I_{x,R}$ and $I_{x,L}$ be the two horizontal terminal currents and $I_{\Sigma}$ the current used for normalization. Combining the electrode geometry, sheet resistance, and readout network into an equivalent horizontal channel weight $m_x(\bm r)$ gives~\cite{HamamatsuPSD}
\begin{equation}\label{eq:psd-readout}
S_x=\frac{I_{x,R}-I_{x,L}}{I_{\Sigma}}=\E_{\mu_{\bm\theta}}[m_x].
\end{equation}
For a two-dimensionally translated Gaussian spot $\bm\theta=(x_0,y_0)^{\mathsf T}$, the source score for vertical displacement is $\ell_{y_0}=2(Y-y_0)/\omega^2$. For an ideal continuous resistive layer, the horizontal channel weight can be approximated as $m_x\simeq\kappa X+b$ within its effective working region. Thus \eqref{eq:covariance-identity} gives
\begin{equation}\label{eq:psd-cross}
\frac{\partial S_x}{\partial y_0}
=\Cov_{\mu_{\bm\theta}}(m_x,\ell_{y_0})
\simeq\frac{2\kappa}{\omega^2}\Cov_{\mu_{\bm\theta}}(X,Y).
\end{equation}
Whether a horizontal PSD channel responds locally to vertical displacement therefore depends on $X$--$Y$ correlation in the actual effective reception distribution; the channel name ``horizontal electrode'' does not guarantee zero response. A finite sensitive area, nonuniform response, or nonideal resistive network can change this covariance. For an actual nonlinear $m_x(\bm r)$, the full first covariance in \eqref{eq:psd-cross} must be retained.

\paragraph{Shack--Hartmann wavefront sensor.}
A Shack--Hartmann sensor converts local wavefront slope into focal-plane spot displacement using a microlens array. For one subaperture, let $\bm\alpha=(\alpha_x,\alpha_y)$ be the local slope, $f$ the microlens focal length, and $G_f(\bm r;\bm\alpha)$ the focal-plane intensity. With a finite detector window $D$ and pixel response $a(\bm r)$, the horizontal centroid is~\cite{Platt2001,Thomas2006}
\begin{equation}\label{eq:sh-centroid}
c_x=\frac{\int_D x\,a(\bm r)G_f(\bm r;\bm\alpha)\,d\bm r}
{\int_D a(\bm r)G_f(\bm r;\bm\alpha)\,d\bm r}
=\E_{\mu_{\bm\alpha}}[X].
\end{equation}
Under small slopes and paraxial conditions, an ideal focal-spot center obeys $\Delta x\simeq f\alpha_x$ and $\Delta y\simeq f\alpha_y$. If the spot of one subaperture is approximately a translated two-dimensional Gaussian of scale $\omega_f$, its source score for $\alpha_y$ is $\ell_{\alpha_y}=2f(Y-f\alpha_y)/\omega_f^2$, so
\begin{equation}\label{eq:sh-cross}
\frac{\partial c_x}{\partial\alpha_y}
=\Cov_{\mu_{\bm\alpha}}(X,\ell_{\alpha_y})
=\frac{2f}{\omega_f^2}\Cov_{\mu_{\bm\alpha}}(X,Y).
\end{equation}
Horizontal and vertical centroid slopes decouple exactly on an ideal infinite uniform focal plane. A finite pixel window, truncation, or nonuniform response can give $\Cov(X,Y)\neq0$ and cross sensitivity in the local centroid readout of a subaperture. Full Shack--Hartmann wavefront recovery also depends on the subaperture-array geometry and global reconstruction matrix~\cite{Southwell1980}, which lie outside this local readout relation.

\paragraph{Pyramid wavefront sensor (PWFS).}
A PWFS splits the incident field with a focal-plane pyramid and forms multiple pupil images at a subsequent detector plane. For a four-pupil configuration, let $P_q(k)$ be the received intensity at pixel position $k$ of pupil image $q$, with $q=1,\ldots,4$. After choosing the pupil numbering and signs, one class of horizontal and vertical normalized channels can be written~\cite{Ragazzoni1996,Fauvarque2016}
\begin{equation}\label{eq:pwfs-readout}
S_i(k)=\frac{\sum_{q=1}^{4}c_{iq}P_q(k)}{Z_k},\qquad
Z_k=\sum_{q=1}^{4}P_q(k),\qquad c_{iq}=\pm1.
\end{equation}
Here $i\in\{x,y\}$ denotes the selected differential channel; the specific $c_{iq}$ depends on pupil numbering. Treat the finite detector regions corresponding to the same $k$ in four pupil images as a joint reception domain $D_k$. Include pixel response in the common reception weight $a(\bm r)$ and the differential signs $c_{iq}$ in $m_{i,k}(\bm r)$, so numerator and denominator contain the same actual response. Parametrize the wavefront by modal coefficients $\bm\beta=(\beta_1,\beta_2,\ldots)$ and define the source score of mode $n$ at the detector plane by $\ell_n(\bm r;\bm\beta)=\partial_{\beta_n}\ln G_{\mathrm{PWFS}}(\bm r;\bm\beta)$. Then
\begin{equation}\label{eq:pwfs-cross}
\frac{\partial S_{i,k}}{\partial\beta_n}
=\Cov_{\mu_{\bm\beta,k}}(m_{i,k},\ell_n).
\end{equation}
Whether a wavefront mode enters a particular PWFS difference channel depends on the projection of its induced detector-plane intensity change onto the channel-combination weight. Finite pupil-image windows, defective pixels, and nonuniform spatial response change this covariance. Unlike the QPD translation example, the PWFS forward field contains diffraction and interference. Thus $G_{\mathrm{PWFS}}$ is the parameter-dependent intensity reaching the detector after the complete optical propagation, not a simply translated Gaussian.

\paragraph{Back-focal-plane (BFP) split detection in optical tweezers.}
In BFP detection, particle displacement changes the interference intensity of scattered and unscattered incident fields in the BFP or a conjugate detector plane. Let this distribution be $G_{\mathrm{BFP}}(\bm r;\bm\theta)$, with $\bm\theta$ containing the particle-position parameters. If left and right detector regions give powers $P_L$ and $P_R$, respectively, a normalized horizontal split channel is~\cite{Gittes1998,Tolic2006}
\begin{equation}\label{eq:bfp-readout}
S_x=\frac{P_R-P_L}{P_R+P_L}
=\E_{\mu_{\bm\theta}}[\sgn X].
\end{equation}
For the particle's vertical coordinate $y_p$, define $\ell_{y_p}=\partial_{y_p}\ln G_{\mathrm{BFP}}(\bm r;\bm\theta)$. Its local contribution to the horizontal split channel is
\begin{equation}\label{eq:bfp-cross}
\frac{\partial S_x}{\partial y_p}
=\Cov_{\mu_{\bm\theta}}(\sgn X,\ell_{y_p}).
\end{equation}
Vertical particle displacement couples into the horizontal BFP channel if its change to the interference pattern has nonzero projection under the left--right sign weight. Neither a Gaussian $G_{\mathrm{BFP}}$ nor a rigid translation of the detector-plane pattern is required. This distinguishes BFP detection from the ideal translated-Gaussian QPD; its particular cross structure must be determined from the actual scattering--interference forward field.

\paragraph{Finite pixel arrays and camera readout.}
A finite pixel array discretizes the detector plane into local integration channels~\cite{Thompson2002,Ober2004}. Let pixel $k$ cover region $P_k$, with received signal $N_k=\int_{P_k}a(\bm r)G(\bm r;\bm\theta)\,d\bm r$ and normalization $Z=\sum_kN_k$ over the selected pixels. With $m_k(\bm r)=\mathbf{1}_{P_k}(\bm r)$, \eqref{eq:channel-expectation} and \eqref{eq:covariance-identity} give the normalized pixel value and its sensitivity:
\begin{equation}\label{eq:pixel-readout}
p_k=\frac{N_k}{Z}=\E_{\mu_{\bm\theta}}[m_k],\qquad
\frac{\partial p_k}{\partial\theta_j}=\Cov_{\mu_{\bm\theta}}(m_k,\ell_j).
\end{equation}
The full pixel vector $(p_1,p_2,\ldots)$ retains multiple projections of a source score onto local spatial indicators. A linear combination $T=\sum_kc_kp_k$ has equivalent weight $m_T=\sum_kc_km_k$ and obeys
\begin{equation}\label{eq:pixel-combination}
T=\E_{\mu_{\bm\theta}}[m_T],\qquad
\frac{\partial T}{\partial\theta_j}=\Cov_{\mu_{\bm\theta}}(m_T,\ell_j).
\end{equation}
Finite regions of interest, split readouts, differential statistics, and discrete centroids based on pixel centers are therefore different coefficient choices for the same array. For example, $c_k=x_k$, where $x_k$ is the horizontal coordinate of pixel center $k$, gives the usual discrete horizontal centroid. More pixels do not automatically increase useful sensitivity to a parameter; what matters is whether new spatial channels give independent projections of its source score. Comparing estimation precision across array designs also requires photon-statistics and readout-noise models beyond these deterministic forward relations~\cite{Mortensen2010}.

The translated Gaussian, sign weights, and fourfold symmetry of the QPD allow the general covariance relation to reduce analytically to $A_{12}$. In the other systems, the detailed cross terms depend on their optical propagation, finite reception, and channel definitions. Throughout, $D$, $a(\bm r)$, and $m_i(\bm r)$ must remain fixed with respect to the parameters considered. If geometry, response, or channel definition itself varies with a parameter, its derivative contributes additional forward-response terms beyond the score--channel covariance.

\section{A QPD example: constructing a two-dimensional inverse from forward structure}\label{sec:inverse}

\subsection{Axis inversion and the cross-residual structure}\label{sec:axis-residual}

We now use the QPD specialization of \cref{sec:theory} to constrain an inverse map. Reflection symmetry removes its cross response on the two coordinate axes, while the response reappears at general off-axis positions. Axis calibration describes the complete one-dimensional nonlinearity along each axis; the two-dimensional correction supplies the joint response that axis data cannot determine.

Let the actual axis responses be $F_x(x_0)=S_x(x_0,0)$ and $F_y(y_0)=S_y(0,y_0)$. Assume that both are continuous and strictly monotone on the selected working intervals, and define the axis-inverted coordinates directly by $\xi=F_x^{-1}(S_x)$ and $\eta=F_y^{-1}(S_y)$. When $y_0=0$ or $x_0=0$, cross response vanishes and correspondingly $\xi=x_0$ or $\eta=y_0$. Thus $(\xi,\eta)$ provide baseline coordinates already corrected for axis nonlinearity.

From the local forward expansion of \cref{sec:covariance},
\begin{equation*}
S_x(x_0,y_0)=F_x(x_0)+A_{12}x_0y_0^2+O(5),
\end{equation*}
a first-order expansion of the axis inverse gives
\begin{equation}\label{eq:inverse-local}
\begin{aligned}
x_0-\xi&=-\frac{A_{12}}{F'_x(x_0)}x_0y_0^2+O(5),\\
y_0-\eta&=-\frac{A_{12}}{F'_y(y_0)}y_0x_0^2+O(5).
\end{aligned}
\end{equation}
Near the origin, $F'_x(0)=F'_y(0)=A_{10}$, so
\begin{equation}\label{eq:inverse-origin}
\begin{aligned}
x_0-\xi&=-\frac{A_{12}}{A_{10}}x_0y_0^2+O(5),\\
y_0-\eta&=-\frac{A_{12}}{A_{10}}y_0x_0^2+O(5).
\end{aligned}
\end{equation}
The residual after axis inversion is not a new single-axis nonlinearity: it is the inverse-map counterpart of the forward cross term. Increasing the accuracy of a one-dimensional axis fit cannot recover orthogonal-coordinate dependence that is identically absent from axis data.

For a finite working domain, define $\Delta_x=x_0-\xi$ and $\Delta_y=y_0-\eta$. Assume that the forward map has a single-valued, sufficiently smooth inverse branch on that domain and that the axis inverses match the actual axis responses. Reflection symmetry requires $\Delta_x$ to be odd in $\xi$ and even in $\eta$, with exchanged parity for $\Delta_y$. Exact axis inversion also requires $\Delta_x(\xi,0)=0$ and $\Delta_y(0,\eta)=0$. If the residual is sufficiently smooth in squared coordinates, its general form is
\begin{equation}\label{eq:residual-general}
\Delta_x=\xi\eta^2\Phi_x(\xi^2,\eta^2),\qquad
\Delta_y=\eta\xi^2\Phi_y(\eta^2,\xi^2).
\end{equation}
The $x_0y_0^2$ and $y_0x_0^2$ terms of \cref{sec:theory} are the lowest-order expansion of this general residual structure, rather than isolated empirical cubics. Higher-order effects of a finite aperture over a larger working domain are contained in the smooth functions $\Phi_x$ and $\Phi_y$.

For the fourfold-symmetric circular QPD considered here, coordinate exchange $x\leftrightarrow y$ further gives $\Phi_x=\Phi_y\equiv\Phi$. The two-dimensional inverse becomes
\begin{equation}\label{eq:acri-inverse}
\begin{aligned}
x_0&=\xi+\xi\eta^2\Phi(\xi^2,\eta^2),\\
y_0&=\eta+\eta\xi^2\Phi(\eta^2,\xi^2).
\end{aligned}
\end{equation}
This form automatically preserves the axis results already fixed by calibration and confines off-axis correction to a function space consistent with detector symmetry. Comparing its lowest-order terms near the origin with \eqref{eq:inverse-origin} gives the analytic anchor
\begin{equation}\label{eq:analytic-anchor}
\Phi(0,0)=-\frac{A_{12}}{A_{10}}.
\end{equation}
Here $\xi,\eta$ retain physical length units, so $A_{10}$, $A_{12}$, and $\Phi$ have dimensions $L^{-1}$, $L^{-3}$, and $L^{-2}$, respectively. If all positions are divided by $R$, the dimensionless coefficients are $a_{10}=RA_{10}$ and $a_{12}=R^3A_{12}$, and the dimensionless correction is $\bar\Phi=R^2\Phi$ with origin value $-a_{12}/a_{10}$.

The axis responses fix the baseline coordinates; reflection symmetry and axis-zero residual fix the cross factors; exchange symmetry makes the two axes share one correction; and the analytic forward coefficient fixes its origin limit. Two-dimensional calibration need determine only the remaining smooth variation of $\Phi$ away from the origin on the finite working domain.

\subsection{Determining the correction from finite calibration data}\label{sec:fit-correction}

We now construct $\Phi$ from finite two-dimensional calibration data under ideal fourfold symmetry. The cross factors, shared-axis relation, and origin anchor $\Phi(0,0)=\Phi_0=-A_{12}/A_{10}$ are already fixed. The task is to represent and determine how $\Phi$ varies away from that point, not to fit the full inverse map again.

Let $|\xi|,|\eta|\leq s$ describe the calibration range in axis-inverted coordinates. Since $\Phi$ depends on squared coordinates, map $\xi^2,\eta^2\in[0,s^2]$ linearly to $[-1,1]$:
\begin{equation}\label{eq:squared-coordinates}
u=\frac{2\xi^2}{s^2}-1,\qquad v=\frac{2\eta^2}{s^2}-1.
\end{equation}
The scale $s$ denotes only the chosen finite calibration domain. Squared coordinates inherit the symmetry constraints directly, while transformation to $[-1,1]$ permits a stable low-order polynomial representation. We use Chebyshev polynomials of the first kind as a numerical basis; this is one implementation of finite-domain smooth approximation, not a condition for the cross-residual structure~\cite{Trefethen2013}.

Let $T_m$ be the degree-$m$ Chebyshev polynomial of the first kind and write $\widetilde\Phi(u,v)=\Phi(s^2(u+1)/2,s^2(v+1)/2)$ to distinguish transformed from physical squared coordinates. An anchored expansion of total degree at most $p$ is
\begin{equation}\label{eq:anchored-chebyshev}
\widetilde\Phi(u,v)=\Phi_0+
\sum_{1\leq m+n\leq p}c_{mn}
\left[T_m(u)T_n(v)-T_m(-1)T_n(-1)\right].
\end{equation}
The physical origin $(\xi,\eta)=(0,0)$ corresponds to $(u,v)=(-1,-1)$, so every fitted basis function vanishes there and \eqref{eq:anchored-chebyshev} maintains $\widetilde\Phi(-1,-1)=\Phi(0,0)=\Phi_0$. Finite calibration data determine only the variation away from the origin; they do not refit $\Phi_0=-A_{12}/A_{10}$.

At two-dimensional calibration position $k$, let the true position be $(x_k,y_k)$ and the readout be $(S_{x,k},S_{y,k})$. Apply the axis inverses of \cref{sec:axis-residual} to define $\xi_k=F_x^{-1}(S_{x,k})$ and $\eta_k=F_y^{-1}(S_{y,k})$. Under fourfold symmetry, one correction function describes horizontal and vertical residuals with exchanged squared coordinates:
\begin{equation}\label{eq:calibration-residuals}
\begin{aligned}
\Delta_{x,k}=x_k-\xi_k&\simeq\xi_k\eta_k^2\widetilde\Phi(u_k,v_k),\\
\Delta_{y,k}=y_k-\eta_k&\simeq\eta_k\xi_k^2\widetilde\Phi(v_k,u_k).
\end{aligned}
\end{equation}
Each off-axis calibration position therefore supplies both horizontal and vertical constraints on the same coefficients $c_{mn}$. This joint fit follows from coordinate-exchange symmetry; it is not an additional empirical assumption. Axis data determine $F_x$ and $F_y$, but the cross factors vanish on the axes, so axis data alone cannot constrain variation of $\Phi$ over the two-dimensional domain. Off-axis samples are required.

We do not fit ratios such as $\Delta_{x,k}/(\xi_k\eta_k^2)$, because the denominator approaches zero near either axis and would amplify discretization and readout errors. Instead, the raw position residuals are fitted jointly:
\begin{equation}\label{eq:fit-objective}
\widehat{\bm c}=\argmin_{\bm c}\sum_k w_k^2\left\{
\left[\Delta_{x,k}-\xi_k\eta_k^2\widetilde\Phi(u_k,v_k;\bm c)\right]^2
+\left[\Delta_{y,k}-\eta_k\xi_k^2\widetilde\Phi(v_k,u_k;\bm c)\right]^2
\right\}.
\end{equation}
Here $w_k>0$ is an optional sample weight, with $w_k=1$ if spatial regions are not reweighted. Because $\widetilde\Phi$ is linear in the unknown $c_{mn}$, \eqref{eq:fit-objective} is a low-dimensional linear least-squares problem.

\begin{figure}[H]\centering
\includegraphics[width=\textwidth,height=.62\textheight,keepaspectratio]{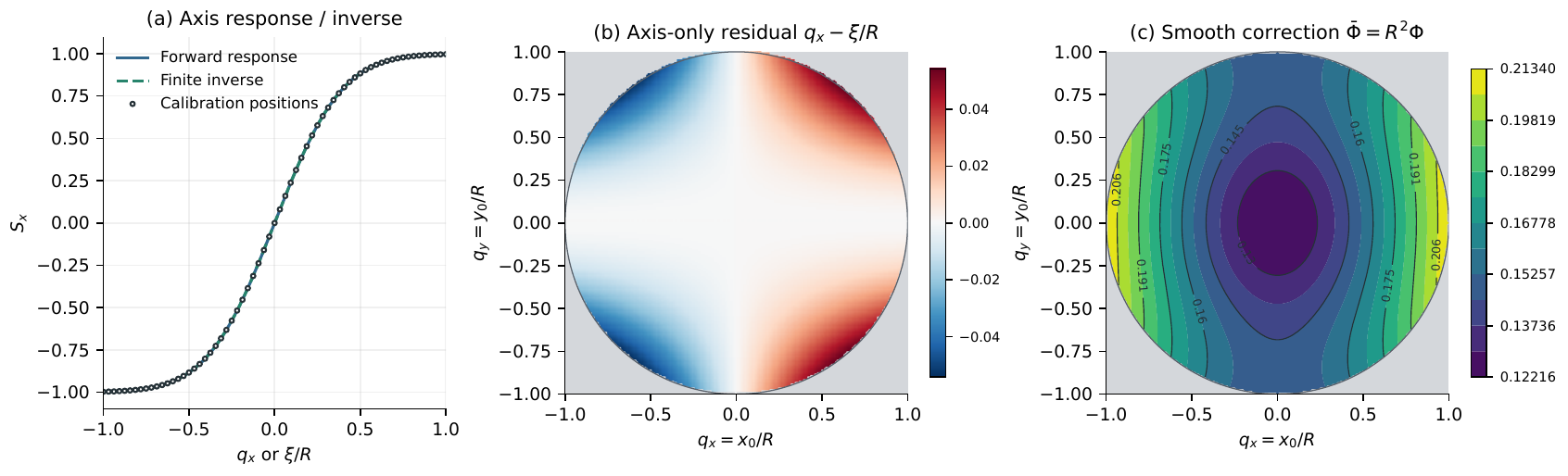}
\caption{Full-disk inverse structure at $\rho=0.45$ and $g/R=0.032$, using 193 physical calibration positions. (a) Forward axis response, the inverse constructed from finite axis calibration, and calibration positions in normalized coordinates; (b) horizontal axis-only residual $q_x-\xi/R$; (c) dimensionless smooth correction $\bar\Phi=R^2\Phi$ after extracting the symmetry-imposed cross factor. Panel (b) is a position residual, whereas panel (c) is a correction function rather than an error map. Finite interpolation leaves a small on-axis error, while the cross-residual correction vanishes exactly on the coordinate axes. Undefined inverse outputs are not extrapolated.}\label{fig:4-1}
\end{figure}

\Cref{fig:4-1} separates these roles. Panel (a) shows finite axis calibration against the forward response; the axis inverse removes that one-dimensional nonlinearity. The residual map in (b) is nevertheless nonzero off axis and changes sign with horizontal position. Panel (c) shows the smooth function left after the symmetry-imposed cross factor is extracted. The axis-zero property refers to the exact axis inverse and cross correction; finite-node interpolation itself retains a discrete approximation error.

The finite-calibration process under ideal fourfold symmetry is thus specified: theory gives the cross factors, shared function, and origin anchor; off-axis data determine the remaining smooth variation of $\Phi$. Expansion degree $p$, calibration scale $s$, and sample layout are choices of numerical representation whose adequacy is tested below.

\subsection{Forward-parameter mismatch and structural preservation under nonideal conditions}\label{sec:mismatch-theory}

The preceding construction uses mirror symmetry, axis-zero residuals, and coordinate-exchange symmetry for a fourfold-symmetric baseline. Real beam widths, apertures, gaps, and spatial responses may depart from nominal values. Two cases must be distinguished. Some deviations preserve the mirror symmetries that determine the cross factors and change only the correction functions and their coefficients. Others break those symmetries and require a larger residual function space.

Consider a detector-axis-aligned elliptical Gaussian:
\begin{equation}\label{eq:ellipse-field}
G(x,y;x_0,y_0)=I_0\exp\!\left[-\frac{(x-x_0)^2}{\omega_x^2}-\frac{(y-y_0)^2}{\omega_y^2}\right].
\end{equation}
Even when $\omega_x\neq\omega_y$, reflections about the $x$ and $y$ axes remain. If each axis inverse uses the actual axis response, the axis-zero residual conditions still hold. The cross-factor structure becomes
\begin{equation}\label{eq:ellipse-residual}
\Delta_x=\xi\eta^2\Phi_x(\xi^2,\eta^2;\bm\lambda),\qquad
\Delta_y=\eta\xi^2\Phi_y(\eta^2,\xi^2;\bm\lambda),
\end{equation}
where $\bm\lambda=(\omega_x/R,\omega_y/R,g/R,\ldots)$ denotes forward-system parameters that preserve these mirror symmetries. The factors $\xi\eta^2$ and $\eta\xi^2$ remain, but $x\leftrightarrow y$ exchange symmetry is generally lost, so usually $\Phi_x\neq\Phi_y$. The shared function $\Phi_x=\Phi_y=\Phi$ is recovered only when beam, reception geometry, and channel responses jointly restore fourfold symmetry. A circular Gaussian thus supplies an additional shared-axis constraint, not a necessary condition for structured two-dimensional inversion.

This distinction also gives two practical calibration levels. If forward parameters such as $R,g,\omega_x,\omega_y$ can be obtained independently, the forward model and axis calibration can be updated directly. For small fabrication or calibration deviations, once the two axis maps have been redetermined, the mirror-symmetry cross factors can be retained and $\Phi_x,\Phi_y$ fitted separately from two-dimensional data. Physical forward parameters and correction-function coefficients play different roles; generally they need not all be estimated simultaneously as free parameters, avoiding unnecessary overparameterization and identifiability problems.

This structural preservation presupposes both mirror symmetries. Asymmetric channel gains, asymmetric occlusion, or an elliptical beam rotated relative to detector coordinates can break them and permit low-order terms excluded from the original model. Refitting the existing $\Phi_x,\Phi_y$ alone then cannot suffice; the residual basis must be extended according to the remaining symmetries. Mirror-symmetry-preserving mismatch chiefly changes the correction functions, whereas symmetry-breaking mismatch can change the allowed structure of those functions.

Deterministic forward-model mismatch must also be separated from random readout noise. Let $(\bm\theta_0,\bm\lambda_0)$ be a nominal working point, with small true shifts $\bm\theta=\bm\theta_0+\delta\bm\theta$ and $\bm\lambda=\bm\lambda_0+\delta\bm\lambda$. Let the observed readout be $\bm S_{\mathrm{obs}}=\bm S(\bm\theta,\bm\lambda)+\bm n$, where $\bm n$ is additive readout noise. Relative to the nominal point, the first-order readout residual is
\begin{equation}\label{eq:mismatch-linearization}
\delta\bm S\simeq J_{\theta}\,\delta\bm\theta
+J_{\lambda}\,\delta\bm\lambda+\bm n.
\end{equation}
The Jacobians $J_{\theta}$ and $J_{\lambda}$ are derivatives with respect to position and forward parameters, both evaluated at the nominal point. Assume $J_{\theta}$ has full column rank and use a consistent local linear inverse at that point, so $\delta\widehat{\bm\theta}\simeq J_{\theta}^{\dagger}\delta\bm S$, where $\dagger$ denotes the Moore--Penrose pseudoinverse. If the forward-parameter shift is ignored, the position-estimation error $\bm e_{\theta}=\delta\widehat{\bm\theta}-\delta\bm\theta$ satisfies
\begin{equation}\label{eq:mismatch-error}
\bm e_{\theta}\simeq J_{\theta}^{\dagger}J_{\lambda}\,\delta\bm\lambda
+J_{\theta}^{\dagger}\bm n.
\end{equation}
The first term is deterministic position bias from departure of the actual forward parameters from the nominal model; updating forward parameters, axis calibration, or correction functions can reduce it. The second term is propagation of random readout noise through the local inverse. It can be reduced by improving signal-to-noise ratio or inverse conditioning, not ``calibrated away'' by adding deterministic coefficients. Both terms depend on the local conditioning of $J_{\theta}$ and connect directly to the $\sigma_{\min}(J_{\theta})$ criterion of \cref{sec:recoverability}.

\section{Numerical simulations and results}\label{sec:simulations}

Controlled numerical tests use the QPD as a specified instance of the finite-domain measurement framework: first to check the forward-sensitivity relations, then to assess the structured inverse they motivate and its applicability limits. The forward readout uses the circular aperture and cross-shaped dead zone of \cref{sec:qpd}, with uniform reception response. Position is normalized as $\bm q=(x_0/R,y_0/R)$ and beam scale as $\rho=\omega/R$, where $\omega$ is the intensity $1/e$ radius and $g$ is the total dead-zone width. The nominal condition is $\rho=0.45$, $g/R=0.032$; $\rho=0.20$ probes smaller-spot behavior but is not used to recommend universal device parameters. Two-dimensional power integration is performed analytically in one coordinate and by 128-point Gauss--Legendre quadrature in the other. Positioning tests cover the entire circular domain $r/R\leq1$. The reported values are model-controlled numerical positioning errors, not measured device accuracy or resolution.

Nominal calibration uses 193 distinct physical positions: 65 on each coordinate axis, sharing the origin, plus 64 off-axis positions. Thus 129 unique positions lie on the two axes and 64 constrain the off-axis cross residual. This layout addresses both the axis inverse and two-dimensional correction; calibration-budget and degree dependence are examined in \cref{fig:5-4}. Calibration readouts are noiseless and position labels exact. Separate validation data are used to select the calibration budget and reference models; the nominal cross-correction total degree is four. Position statistics use a separate set of 8192 points uniformly distributed in disk area; each directional profile uses 401 equally spaced radial points. The normalized Euclidean position residual and its root mean square are
\begin{equation}\label{eq:error-metric}
e_k=\|\widehat{\bm q}_k-\bm q_k\|_2,\qquad
\RMSE=\sqrt{\frac{1}{N_v}\sum_{k\in V}e_k^2}.
\end{equation}
Here $V$ is the set of valid outputs and $N_v$ its size; P95 is the 95th percentile of errors within that set. The error is not divided again by the current position radius. Undefined or numerically saturated readouts are invalid. Readouts are neither clipped nor extrapolated beyond the axis-inverse tables; finite estimates outside the unit disk remain in the error calculation. When invalid outputs occur, RMSE and P95 are conditional statistics and must be reported together with valid coverage.

\subsection{Numerical checks of the forward-sensitivity relations}\label{sec:forward-tests}

\begin{figure}[H]\centering
\includegraphics[width=\textwidth,height=.68\textheight,keepaspectratio]{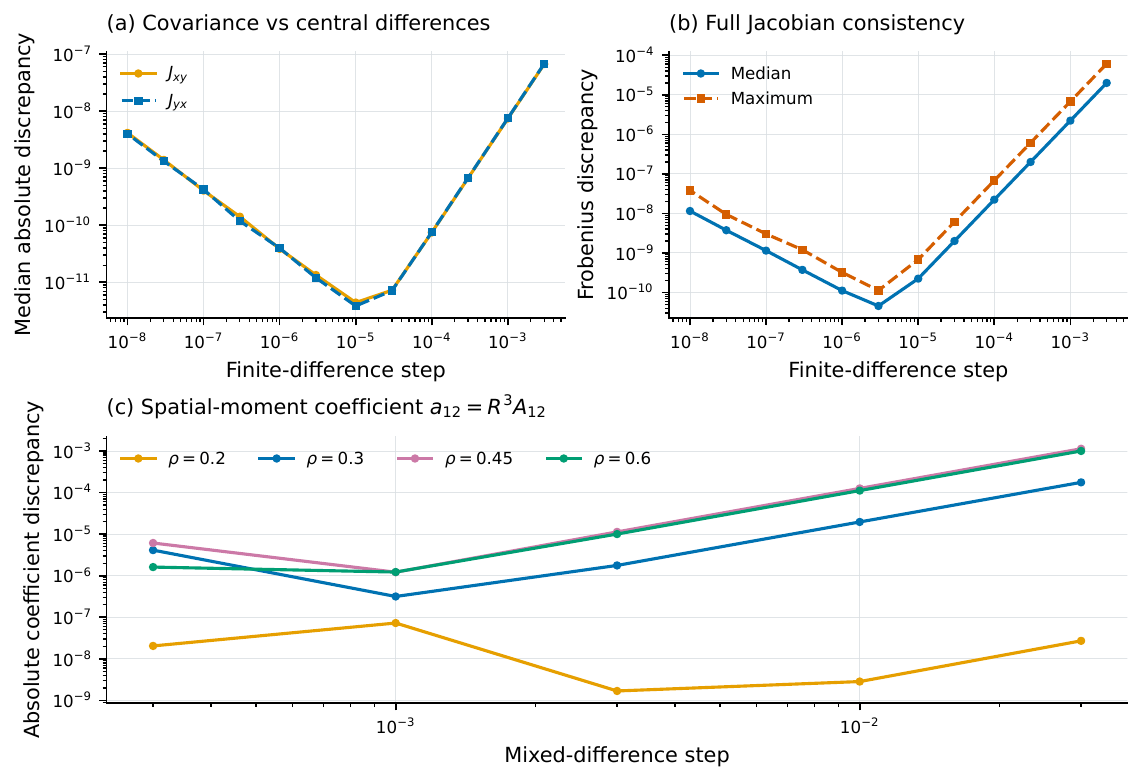}
\caption{Numerical consistency of the forward-sensitivity relations. (a) Median absolute discrepancies of the cross derivatives. (b) Median and maximum Frobenius discrepancies of the full Jacobian with respect to $\bm q$. (c) Absolute discrepancies of the dimensionless coefficient $a_{12}$ at four beam sizes. The total gap $g/R=0.032$ is fixed. Panels (a,b) use the same 260 positions and include the complete finite-difference step sweep.}\label{fig:5-1}
\end{figure}

To test the local-sensitivity expression of \cref{sec:covariance}, we fix forward parameters and compare the covariance expression with central finite differences of the complete normalized readout at the same 260 positions. Finite differences do not call analytic derivatives; both calculations form $J_q$ with respect to normalized position. \Cref{fig:5-1}(a,b) checks the cross derivatives and full Jacobian, respectively. The discrepancy first falls as the finite-difference step decreases and then rises under floating-point cancellation. Near $h=3\times10^{-6}$, the median Frobenius discrepancy is $4.54\times10^{-11}$. Similar discrepancy statistics for the two cross derivatives do not mean that their values coincide at each position.

\Cref{fig:5-1}(c) uses third-order finite differences of the complete forward readout to check the dimensionless cross coefficient $a_{12}=R^3A_{12}$ for the same dead-zone width and four beam scales. For the nominal beam at $h=3\times10^{-3}$, the relative discrepancy is $3.60\times10^{-5}$. The small-beam coefficient approaches zero, so the panel consistently reports absolute discrepancies. Independent two-dimensional integration and quadrature-order checks of the forward calculation, together with the separate no-gap circular-aperture comparison with the published Eq.~(37), are given in the \supp. That no-gap case is distinct from the gapped conditions of \cref{fig:5-1}(c). These tests support the local derivative and mixed-coefficient expressions; they do not prove global recoverability.

\subsection{ACRI positioning and finite-calibration performance}\label{sec:acri-performance}

The nominal ACRI implementation is fixed as follows. Axis samples form inverse functions through odd-symmetric piecewise cubic Hermite interpolation~\cite{FritschCarlson1980,FritschButland1984}, using the interpolation coordinate $t=\rho\operatorname{erf}^{-1}(S)$. The cross correction in normalized squared coordinates has total degree $p=4$ and shares 14 residual coefficients between axes, with analytic anchor $-a_{12}/a_{10}$. The joint fit of \eqref{eq:fit-objective} uses $w_k=(1+r_k^2/R^2)^{-1/2}$. Online positioning performs only interpolation and basis evaluation; it does not call forward quadrature or an iterative root solver. The 14 coefficients exclude both axis interpolation tables and are therefore not the total number of calibration degrees of freedom.

\begin{figure}[H]\centering
\includegraphics[width=\textwidth,height=.72\textheight,keepaspectratio]{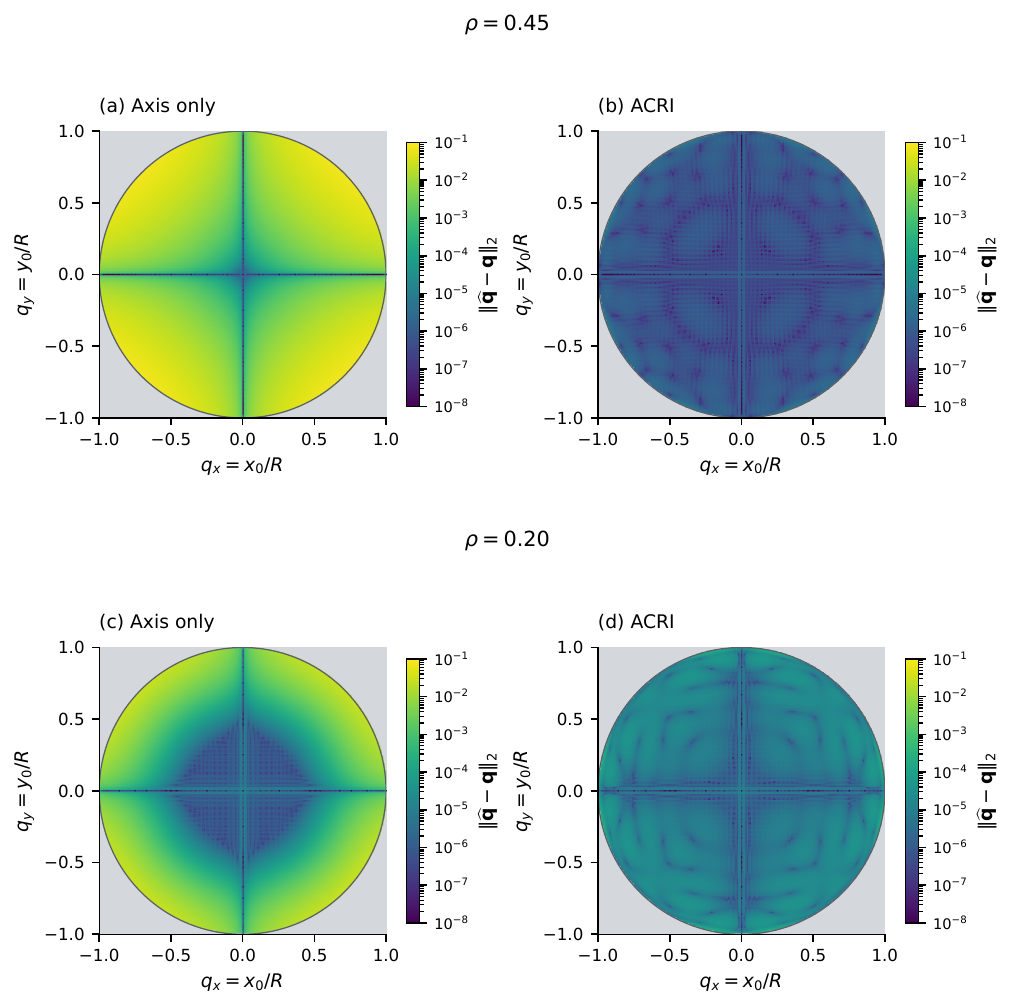}
\caption{Controlled ablation of cross-residual correction at $\rho=0.45$ (a,b) and $\rho=0.20$ (c,d). (a,c) Axis-only inversion. (b,d) ACRI using exactly the same finite axis calibration. All panels use the same full-disk grid and logarithmic color range for the normalized Euclidean position residual \(|\hat{\mathbf q}-\mathbf q|_2\). Values outside the displayed color limits saturate; the lower color limit is not an accuracy floor. Invalid outputs are gray.}\label{fig:5-2}
\end{figure}

To isolate the contribution of cross correction, we keep exactly the same axis calibration and remove only the two-dimensional residual term. \Cref{fig:5-2} uses the same full-disk grid and color range for both beam sizes. Low-error regions for axis-only inversion remain near the coordinate axes, while general off-axis positions show substantial bias. ACRI markedly reduces off-axis bias, and its correction remains identically zero on the axes, as predicted in \cref{sec:qpd,sec:theory}.

Both beam sizes benefit from cross correction, but errors from finite nodes and finite correction degree vary with beam scale and position. Larger residuals remain for the smaller spot and near the aperture edge. Nominal high accuracy must therefore be stated together with beam parameters, calibration layout, and the full working domain.

Under the same 193 physical calibration positions and common test protocol, we compare ACRI with four one-dimensional inverse families and an unstructured two-dimensional inverse. The one-dimensional models in \cref{tab:5-1} use inverse-erf, scale-calibrated erf, odd-polynomial, and Boltzmann-type logarithmic forms. These choices represent function families used in axis-wise QPD calibration, including erf-based compensation, polynomial fitting, and Boltzmann-based correction~\cite{Wu2015,Wang2020,ZhangW2019}. Their parameters are calibrated under the present protocol; they are not paper-by-paper reproductions of published composite algorithms. Absorbing geometric effects through a scale parameter is related to effective-radius modeling~\cite{Li2018}; here the Gaussian inverse erf uses one shared scale parameter. The unstructured two-dimensional comparator directly uses both readout channels without imposing this paper's symmetry or analytic structure.

\begin{table}[H]\centering\footnotesize
\caption{Definitions of the comparison methods.}\label{tab:5-1}
\begin{tabularx}{\textwidth}{@{}>{\raggedright\arraybackslash}p{1.05in}>{\raggedright\arraybackslash}p{1.65in}>{\raggedright\arraybackslash}p{1.45in}>{\raggedright\arraybackslash}X@{}}
\toprule
Method & Inverse form & Calibration & Structural feature\\
\midrule
Inverse-erf & $q_j^{\rm erf}=\rho\operatorname{erf}^{-1}(S_j)$ & None & Axis-wise analytic inverse for an ideal infinite-plane Gaussian\\
Scale-calibrated erf & $q_j^{\rm scale}=\kappa\rho\operatorname{erf}^{-1}(S_j)$ & One shared scale $\kappa$ & Retains the erf shape; changes only overall scale\\
1D polynomial & $q_j^{\rm poly}=\sum_{k=1}^{11}c_kT_{2k-1}(S_j)$ & Common calibration data; validation-set degree selection & Odd-symmetric axis fit with no mixed-channel terms\\
Boltzmann-type logarithmic inverse & $q_j^{\rm B}=b_j\ln[(1+S_j)/(1-S_j)]$ & One $b_j$ per axis & Centrally symmetric one-dimensional parametric inverse\\
ACRI & Axis inverse plus cross-residual correction & Axis interpolation and 14 two-dimensional residual coefficients & Explicit orthogonal-coordinate dependence\\
Unstructured 2D & $q_j=F_j(S_x,S_y)$; total-degree bivariate Chebyshev expansion & Same 193 points; validation-selected degree; 110/42 coefficients & Independent outputs; no imposed symmetry, axis-zero residual, or analytic anchor\\
\bottomrule
\end{tabularx}
\end{table}

All one-dimensional fitted references use position least squares on the common calibration set. The polynomial contains only odd terms; candidates, including a monotonicity-constrained version, are selected on 4096 independent validation points. Both beam sizes select degree 21, with the constrained version used for the smaller spot. The unstructured two-dimensional comparator takes $q_j=\sum_{m+n\leq d}b_{jmn}T_m(S_x)T_n(S_y)$ with independently fitted axis coefficients, including constant, odd, even, and mixed terms. It does not impose axis-zero residuals, a shared function, or an analytic anchor. Total degree is chosen from 1--15 by RMSE on the same validation set, discarding column-rank-deficient configurations; the final test set is not used for selection. Degrees 9 and 5 are selected for the two beam sizes, giving 110 and 42 coefficients across both outputs. Both use the same 193 calibration positions. This comparison assesses the specified function families and calibration layout, not the best possible performance of every two-dimensional approximation method.

\Cref{tab:5-2} gives the full-disk test results; every method has valid output at all 8192 positions. ACRI's normalized RMSE is $1.054\times10^{-6}$ for the nominal beam and $2.347\times10^{-5}$ for the smaller beam. The difference between axis-only inversion and ACRI isolates the cross correction. The unstructured two-dimensional comparator can express cross dependence but does not reach the same accuracy for this calibration layout and finite polynomial space. The result supports the use of forward constraints to organize finite calibration data; it does not attribute the advantage solely to whether mixed terms are present. Degree-selection curves and directional residuals are supplied in the \supp.

\begin{table}[H]\centering\small
\caption{Full-disk position errors under the common finite-calibration protocol.}\label{tab:5-2}
\begin{tabular}{@{}lrrrr@{}}
\toprule
Method & RMSE $\rho=0.45$ & P95 $\rho=0.45$ & RMSE $\rho=0.20$ & P95 $\rho=0.20$\\
\midrule
Inverse-erf & $3.679\times10^{-2}$ & $7.253\times10^{-2}$ & $1.658\times10^{-2}$ & $2.215\times10^{-2}$\\
Scale-calibrated erf & $1.912\times10^{-2}$ & $3.816\times10^{-2}$ & $1.003\times10^{-2}$ & $1.539\times10^{-2}$\\
1D polynomial & $1.969\times10^{-2}$ & $4.424\times10^{-2}$ & $1.528\times10^{-1}$ & $2.568\times10^{-1}$\\
Boltzmann & $5.386\times10^{-2}$ & $8.154\times10^{-2}$ & $1.293\times10^{-1}$ & $1.723\times10^{-1}$\\
ACRI & $1.054\times10^{-6}$ & $1.800\times10^{-6}$ & $2.347\times10^{-5}$ & $3.665\times10^{-5}$\\
Axis only & $2.528\times10^{-2}$ & $5.539\times10^{-2}$ & $6.838\times10^{-3}$ & $1.741\times10^{-2}$\\
Unstructured 2D & $3.146\times10^{-2}$ & $5.734\times10^{-2}$ & $2.051\times10^{-1}$ & $3.356\times10^{-1}$\\
\bottomrule
\end{tabular}
\end{table}

\begin{figure}[H]\centering
\includegraphics[width=\textwidth,height=.76\textheight,keepaspectratio]{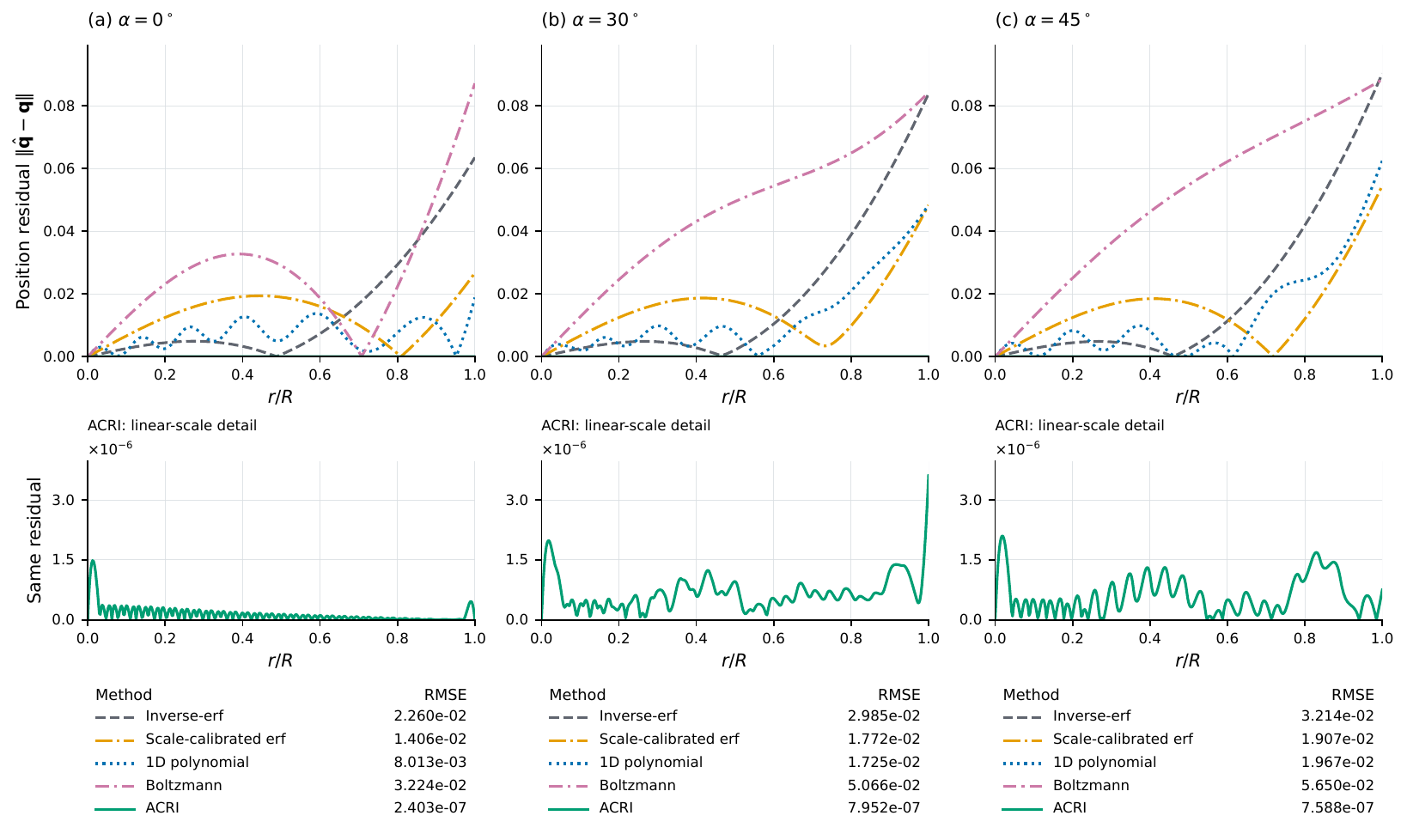}
\caption{Pointwise normalized position residuals in three directions at $\rho=0.45$. Every fitted method receives the same 193 calibration positions; inverse-erf is an unfitted reference. Each direction uses 401 test positions. Linear axes retain all samples, including exact zeros. Lower strips show the same ACRI residuals at finer linear scale. Each legend gives RMSE along that radial profile, rather than the area-uniform full-disk RMSE.}\label{fig:5-3}
\end{figure}

\Cref{fig:5-3} further compares pointwise residuals from the origin to the aperture edge along $\alpha=0^\circ$, $30^\circ$, and $45^\circ$ at $\rho=0.45$, $g/R=0.032$. The $0^\circ$ direction is exactly axial, where the theoretical cross residual vanishes, and mainly tests each method's description of finite-aperture single-axis nonlinearity. The $30^\circ$ direction is generally off axis; at $45^\circ$ both coordinates change equally, a representative joint displacement in a fourfold-symmetric system. The vertical axis is the normalized Euclidean residual of \eqref{eq:error-metric}; the fine-scale lower panels show the same ACRI data. Legend RMSE is computed from 401 equally spaced radial positions at each angle and differs from the disk-area-uniform statistic in \cref{tab:5-2}.

The local minima in \cref{fig:5-3} can arise when signed bias crosses zero, without a periodic oscillation in the forward response. Small axial ripples in ACRI retain finite-node interpolation error; off-axis curves also contain finite-degree approximation residuals of the correction function. These variations have not been smoothed or removed. Directional profiles for the smaller beam are given in the \supp.

\begin{figure}[H]\centering
\includegraphics[width=\textwidth,height=.70\textheight,keepaspectratio]{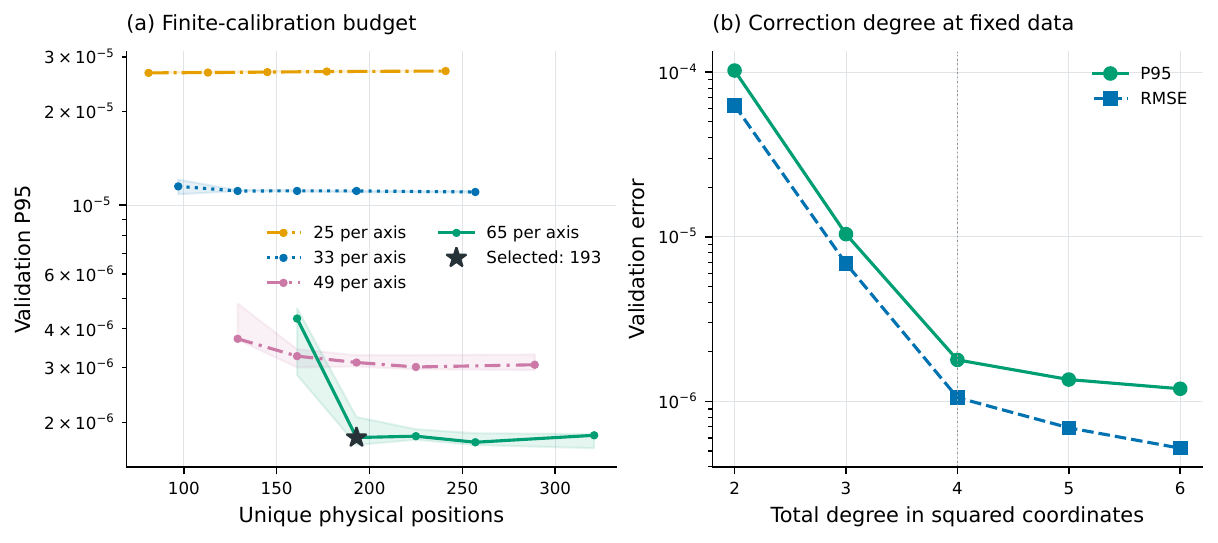}
\caption{ACRI calibration-budget and correction-degree dependence at $\rho=0.45$. (a) The correction degree and validation set are fixed while axis and off-axis sample counts vary. Lines show the median P95 over three off-axis designs; shaded bands span their minimum and maximum and are not confidence intervals. (b) The same 193 physical positions are retained while total degree in squared coordinates changes. Neither comparison uses final-test points.}\label{fig:5-4}
\end{figure}

To assess calibration budget, we fix the forward model, degree-four correction, and independent validation set, then vary axis-node and off-axis-sample counts. Each configuration uses three off-axis sampling realizations. In \cref{fig:5-4}(a), each curve keeps its axis-node density fixed. With sparse axis nodes, adding off-axis positions cannot remove the error plateau caused by axis interpolation. After increasing axis density, additional off-axis sampling reduces error further. The 193-position configuration lies in a region of diminishing returns, but is not claimed to be a globally minimal calibration set.

\Cref{fig:5-4}(b) holds the same 193 positions and changes only the cross-correction degree. Validation error falls markedly from degree two to four; higher degree gives further but smaller gains. Degree four is therefore used as a low-dimensional nominal setting, not chosen to minimize final-test error. The \supp\ also considers perturbations of calibration readouts and position labels: performance of the nominal beam degrades continuously, while the smaller beam is more sensitive to calibration noise because axis readouts approach saturation and nodes compress.

\subsection{Sensitivity to nonideal factors and applicability limits}\label{sec:nonideal}

\begin{figure}[H]\centering
\includegraphics[width=\textwidth,height=.72\textheight,keepaspectratio]{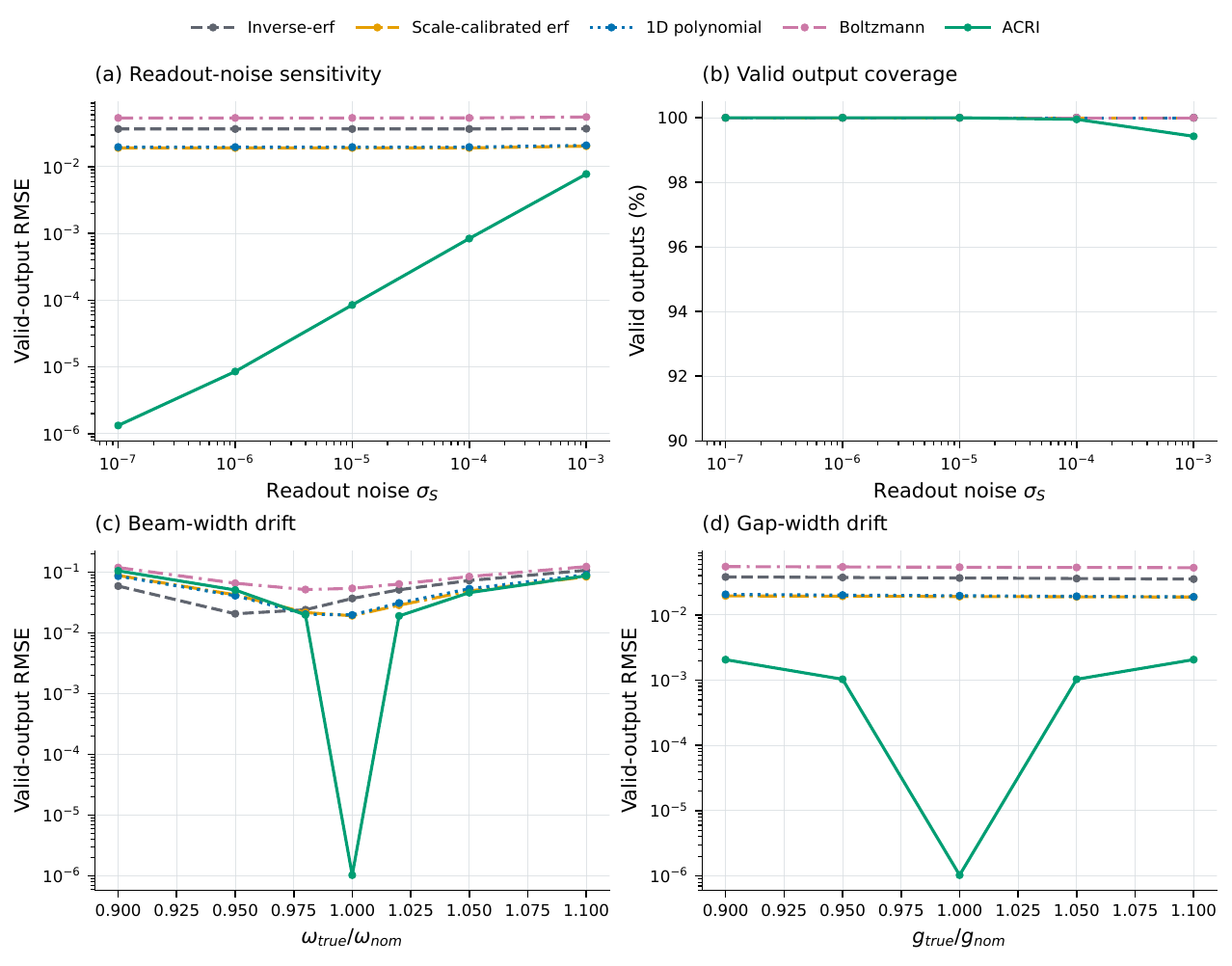}
\caption{Sensitivity under fixed nominal calibration at $\rho=0.45$ and $g/R=0.032$. (a,b) Shared additive readout-noise realizations at 2048 positions with eight repetitions. (c,d) Noiseless changes to forward parameters at the same positions; all inverse coefficients remain fixed. RMSE is conditional on valid outputs. For ACRI, valid coverage under beam-width drift ranges from 90.14\% to 100\%; coverage under gap drift is 100\%. Finite estimates outside the unit disk are retained.}\label{fig:5-5}
\end{figure}

To distinguish random noise from deterministic mismatch, \cref{fig:5-5} first fixes every nominal inverse without recalibration. Readout noise is added only to the test channels $S_x,S_y$ as independent zero-mean Gaussian perturbations. All methods use the same 2048 positions and eight noise realizations. In \cref{fig:5-5}(a), ACRI's conditional RMSE grows approximately with noise standard deviation over the examined range: its low deterministic residual allows readout noise to dominate earlier.

The four one-dimensional references change little over this noise range, but this should not be interpreted as superior noise robustness. Their noiseless deterministic model bias is already of order $10^{-2}$, so readout perturbations below that floor are initially masked by the existing bias. ACRI reveals the noise limit sooner precisely because its nominal residual is lower. \Cref{fig:5-5}(b) additionally reports valid-output fraction. Axis interpolation here is not extrapolated beyond its calibrated readout interval, so larger noise can push some observations outside the inverse domain. At $\sigma_S=10^{-3}$, ACRI still has 99.43\% valid output. Full coverage variation is in the \supp.

\Cref{fig:5-5}(c,d) adds no random noise. Only the true parameters generating the readout change; all inverse coefficients remain nominal, so these panels test deterministic calibration mismatch. A beam-width departure of about 2\% from nominal already raises ACRI's normalized RMSE to about $2\times10^{-2}$. Its small nominal residual does not automatically cancel a systematic scale drift. Decreasing beam width can also reduce valid coverage, whose range is reported in the caption; conditional RMSE alone therefore does not characterize robustness. Pointwise coverage is given in the \supp. A similar relative change in dead-zone width produces a smaller error in \cref{fig:5-5}(d), though still well above the nominal numerical residual. This sensitivity ranking applies only to the present aperture, beam, and working domain.

\FloatBarrier

\begin{figure}[H]\centering
\includegraphics[width=\textwidth,height=.72\textheight,keepaspectratio]{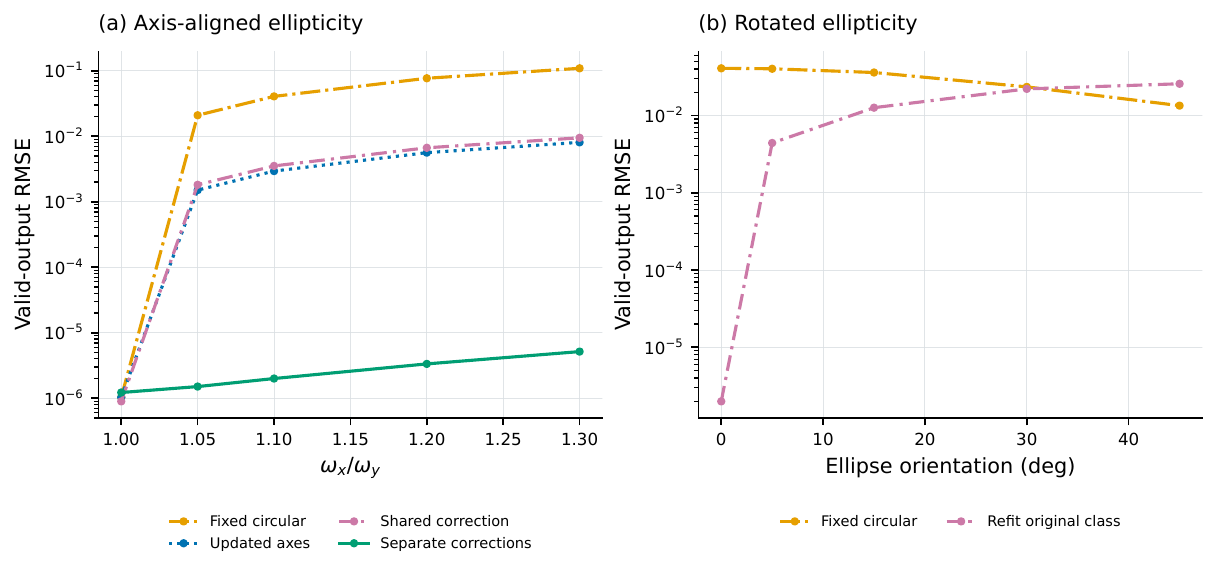}
\caption{Recalibration under elliptical illumination at fixed geometric-mean beam radius. (a) Axis-aligned ellipses: fixed circular calibration, updated axes only, refitted shared correction, and refitted separate corrections. Refits use the same physical positions; unanchored shared and separate degree-four corrections contain 15 and 30 coefficients. (b) Fixed aspect ratio 1.1 with varying orientation: circular calibration versus refitting the original parity-constrained class. RMSE uses valid outputs. Fixed-circular coverage ranges from 93.85\% to 100\% in (a) and from 98.34\% to 100\% in (b); recalibrated models have full coverage.}\label{fig:5-6}
\end{figure}

The fixed-calibration tests above assess performance after parameter drift. A separate question is whether ACRI's function class from \cref{sec:inverse} remains applicable after recalibration. \Cref{fig:5-6} addresses this question by changing system symmetry through elliptical illumination and comparing recalibration strategies.

First, keep geometric-mean beam scale $\sqrt{\omega_x\omega_y}/R=0.45$ and the aperture and gap fixed, while increasing the aspect ratio $\omega_x/\omega_y$ of an ellipse aligned with detector axes. Reflections $x\mapsto-x$ and $y\mapsto-y$ remain, but $x\leftrightarrow y$ exchange symmetry is lost when $\omega_x\neq\omega_y$. \Cref{fig:5-6}(a) compares four treatments: Fixed circular retains the complete circular-beam calibration; Updated axes redetermines only the two axis inverses; Shared correction updates the axes but still forces a common correction function; Separate corrections fits $\Phi_x$ and $\Phi_y$ separately while retaining the mirror-symmetry cross factors. All three recalibrated treatments use the same 193 physical positions as the nominal model.

Updating only the axes removes most of the one-dimensional response change caused by ellipticity in \cref{fig:5-6}(a), but forcing a shared correction does not restore the original accuracy. Releasing the shared constraint reduces the residual again to order $10^{-6}$. This agrees with \cref{sec:mismatch-theory}: the cross factors survive when both mirror symmetries remain, while loss of exchange symmetry requires separate functions for the axes.

\Cref{fig:5-6}(b) fixes aspect ratio at 1.1, changes orientation, and recalibrates the original parity-constrained model at each angle using the same positions. After rotation, even refitting every coefficient cannot recover the low residual of the axis-aligned case. A generally rotated ellipse introduces $xy$ coupling in detector coordinates and breaks the separate $x$ and $y$ reflection symmetries. Terms forbidden by the original function class then become allowed, and refitting its existing coefficients cannot supply those missing terms. The fixed circular model has lower error at some rotation angles, but that does not mean its structure becomes valid again: full-domain conditional RMSE depends on the spatial distribution of bias and on the valid set. The conclusion is limited to this point: after mirror-symmetry breaking, refitting within the original function class is insufficient to restore axis-aligned performance, and the admissible residual space must be expanded.

\FloatBarrier

These QPD tests illustrate how the general forward representation constrains the inverse function class. ACRI's applicability depends on the forward state, calibration conditions, and remaining symmetries: loss of exchange symmetry requires separate correction functions, while loss of mirror symmetry requires a broader residual space. Device accuracy and resolution must still be established experimentally, with an uncertainty budget~\cite{JCGM2008} accounting for beam stability, stage positioning, channel gain, photon statistics, and electronic noise.

\section{Conclusion}\label{sec:conclusion}

Finite-domain normalized spatial measurements can be represented as channel-weight expectations under an effective measure determined by the incident distribution, spatial response, and receiving support. With these measurement operators fixed, the covariance between channel weight and source score gives local sensitivity. This representation separates three questions: whether each channel depends on only one parameter, whether parameters mix in the readout, and whether the joint map locally recovers them. Nonzero cross sensitivity does not by itself imply information loss; local recoverability depends on the joint Jacobian and does not establish global injectivity.

The QPD provides an analytical example. When the incident distribution and complete effective reception structure do not retain compatible product forms, normalization can preserve orthogonal-coordinate dependence. The mixed coefficient quantifies its lowest-order contribution under the stated symmetries. Coupling can vanish on the axes, so accurate axis calibration alone does not establish separability of the two-dimensional map. ACRI uses this forward structure to constrain the off-axis residual after axis inversion, demonstrating how a measurement model can organize a finite-data inverse.

The numerical results support this construction for the specified QPD conditions; performance depends on beam scale, operating position, calibration state, and retained symmetries. The general representation also applies to other fixed-operator measurements with finite reception, spatial weighting, and normalized readout, whereas the particular ACRI residual factors rely on the QPD symmetries. The resulting route connects reception structure, parameter coupling, local recoverability, and constrained inversion without asserting a universal device accuracy.

\bibliographystyle{unsrt}
\begingroup
\small
\interlinepenalty=10000
\bibliography{references}
\endgroup
\end{document}